%% file: main.tex
\documentclass[runningheads]{llncs}

\input{usepackages.tex}

\input{packages/macros.tex}

\input{packages/macros-declare.tex}

\begin{document}

\title{Reactive Synthesis for DECLARE via symbolic automata}
\author{%
  Luca Geatti \inst{1} \and 
Marco Montali \inst{2} \and 
Andrey Rivkin \inst{2} 
}
\authorrunning{L. Geatti et al.}
\institute{
  University of Udine,
  Via delle Scienze, 206,
  Italy - 33100, Udine \\
  \email{luca.geatti@uniud.it}
  \and
  Free University of Bozen-Bolzano, 
  Piazza Università 1,
  Italy - 39100, Bozen-Bolzano \\
  \email{\{montali,rivkin\}@inf.unibz.it}
}

\maketitle

\begin{abstract}
  Given a specification of linear-time temporal logic interpreted over
  finite traces (\LTLf), the reactive synthesis problem asks to find
  a finitely-representable, terminating controller that reacts to the
  uncontrollable actions of an environment in order to enforce a desired system
  specification. 
  In this paper we study, for the first time, the reactive synthesis problem for {\DECLARE}  --  a fragment of \LTLf extensively used both in theory and practice for specifying declarative, constraint-based business processes. We provide a threefold contribution. First, we give a na\"ive, doubly exponential time synthesis algorithm for this problem. 
  Second, we show how an arbitrary \DECLARE specification can be compactly
  encoded into an equivalent pure past one in \LTLf, and we exploit this to define an optimized, singly exponential time algorithm for \DECLARE synthesis.
  Third, we derive a symbolic version of this algorithm, by introducing a novel translation of pure-past temporal
  formulas into symbolic deterministic finite automata.
\end{abstract}

\input{sections/1.intro.tex}

\input{sections/2.background.tex}
\input{sections/3.declaresynthesis.tex}

\input{sections/4.explicit.tex}
\input{sections/5.symbolic.tex}
\input{sections/6.conclusions.tex}

\bibliographystyle{splncs04}
\bibliography{main}

\end{document}

%% file: usepackages.tex
\usepackage[utf8]{inputenc}
\usepackage[T1]{fontenc}
\usepackage[inline]{enumitem}
\usepackage[all]{foreign}
\usepackage{amsmath,amssymb}
\usepackage{thm-restate}
\usepackage{subfigure}
\usepackage{lineno,hyperref}
\usepackage[capitalise]{cleveref}
\usepackage{xspace}
\usepackage{thmtools}
\usepackage{mathtools}
\usepackage{autonum}
\usepackage[all]{foreign}
\usepackage{xparse}
\usepackage[table]{xcolor}
\usepackage{forest}
\usepackage{makecell}
\usepackage{multicol}
\usepackage{rotating}
\usepackage{marginnote}
\usepackage{tikz}
\usepackage{algorithm}
\usepackage{algpseudocode}
\usepackage{todonotes}
\usepackage{booktabs}
\usepackage{lipsum}
\usepackage{packages/commands} 
\usepackage{packages/graphics} 
\usepackage{packages/logic}    
\usepackage{packages/symbols}  
\usepackage{breakcites}

\usepackage{graphicx}
\usepackage{xspace}
\usepackage{tikz,subfigure}
\usetikzlibrary{arrows,shapes,shapes.multipart,snakes,automata,backgrounds,petri,positioning,shadows,matrix,decorations.pathmorphing, decorations.pathreplacing, decorations.markings, fit,positioning,calc,backgrounds,shapes.misc,arrows.meta,fit}

%% file: packages/macros.tex
\newcommand{\mypar}[1]{\smallskip\noindent\textbf{#1.}~}

\makeatletter
\g@addto@macro\normalsize{%
\setlength{\abovecaptionskip}{0pt}
\setlength{\belowcaptionskip}{-10pt}
\setlength\abovedisplayskip{3pt}
\setlength\belowdisplayskip{3pt}
\setlength\abovedisplayshortskip{3pt}
\setlength\belowdisplayshortskip{3pt}
}
\makeatother

\newcommand{\str}{\mathtt{string}\xspace}
\newcommand{\bool}{\mathtt{bool}\xspace}

\newcommand{\activity}[1]{\texttt{#1}}

\definecolor{deepblue}{HTML}{0C3B80}
\definecolor{deepgreen}{HTML}{2EA601}
\definecolor{lightOrange}{HTML}{FFA03C}
\definecolor{darkOrange}{HTML}{F1800A}
\definecolor{lightBlue}{HTML}{0174CD}
\definecolor{greenF}{HTML}{2CBB5C}
\definecolor{cyan}{HTML}{86A6D5}
\definecolor{darkred}{HTML}{8B0000}

\def\DTZU {2ex}

\tikzstyle{taskfg} = [
  text=deepblue,
]

\tikzstyle{taskbg} = [
  fill=cyan!50,
]

\tikzstyle{taskline} = [
  draw=deepblue,
]

\tikzstyle{taskstyle} = [
  ultra thick,
  taskfg,
  taskbg,
  taskline
]

\tikzstyle{task} = [
  rectangle,
  minimum width=12mm,
  minimum height=10mm,
  taskstyle
]

\tikzstyle{smalltask} = [
  rectangle,
  minimum width=8mm,
  minimum height=6mm,
  taskstyle,
  very thick
]

\tikzstyle{constraint} = [
  ultra thick,
  taskline,
  taskbg
]

\tikzstyle{response} = [
  constraint,
  *-triangle 60
]

\tikzstyle{precedence} = [
  constraint,
  *triangle 60-
]
\tikzstyle{succession} = [
  constraint,
  *triangle 60-*
]

\tikzstyle{respondedexistence} = [
  constraint,
  *-
]

\tikzstyle{coexistence} = [
  constraint,
  *-*
]

\tikzstyle{negationconstraint} = [
  constraint,
  postaction={decorate,decoration={markings,
   mark=at position .5 with {\arrow[xshift=0.15*\DTZU]{Bar[width=1.5*\DTZU]}},
   mark=at position .5 with {\arrow[xshift=-0.15*\DTZU]{Bar[width=1.5*\DTZU]}}
  }}
]

\tikzstyle{notcoexistence} = [
  negationconstraint,
  coexistence,
]

\tikzstyle{negationresponse} = [
  constraint,
  response,
  postaction={decorate,decoration={markings,
   mark=at position .5 with {\arrow[xshift=0.15*\DTZU]{Bar[width=1.5*\DTZU]}},
   mark=at position .5 with {\arrow[xshift=-0.15*\DTZU]{Bar[width=1.5*\DTZU]}}
  }}
]

\tikzstyle{negationsuccession} = [
  constraint,
  succession,
  postaction={decorate,decoration={markings,
   mark=at position .65 with {\arrow[xshift=0.15*\DTZU]{Bar[width=1.5*\DTZU]}},
   mark=at position .65 with {\arrow[xshift=-0.15*\DTZU]{Bar[width=1.5*\DTZU]}}
  }}
]

\tikzstyle{exclusivechoice} = [
  constraint,
  -,
  postaction={decorate,decoration={markings,
   mark=at position .5 with {\arrow[xshift=0.5*\DTZU]{Diamond[width=1*\DTZU]}} 
  }}
]

\tikzstyle{choice} = [
  constraint,
  -,
  postaction={decorate,decoration={markings,
   mark=at position .5 with {\arrow[xshift=0.5*\DTZU]{Diamond[width=1*\DTZU]}},
    mark=at position .5 with {\arrow[xshift=0.28*\DTZU,white,scale=.6]{Diamond[width=1*\DTZU]}} 
  }}
]

\tikzstyle{circ} = [
  solid,
  text=black
]

\tikzstyle{timeline} = [
  -,
  thin,
  ]
  
\tikzstyle{snapshot} = [
  -,
  thin,
  densely dotted,
  ]

\tikzstyle{objnode} = [
  inner sep=2pt,
  circle,
  ultra thick,
]

\tikzstyle{tobj} = [
  objnode,
  taskfg,
  taskbg,
  taskline
]

\tikzstyle{cobj} = [
  objnode,
  classfg,
  classbg,
  classline
]

\tikzstyle{link} = [
  solid,
  -angle 60
]

\tikzstyle{place}=[circle,thick,draw=black,fill=white,minimum size=7mm,font=\fontsize{9}{144}\selectfont]
\tikzstyle{transition}=[rectangle,thick,draw=black,fill=gray!20,minimum size=7mm]

%% file: packages/macros-declare.tex
\renewcommand{\activity}[1]{\texttt{#1}}

\definecolor{deepblue}{HTML}{0C3B80}
\definecolor{deepgreen}{HTML}{2EA601}
\definecolor{lightOrange}{HTML}{FFA03C}
\definecolor{darkOrange}{HTML}{F1800A}
\definecolor{lightBlue}{HTML}{0174CD}
\definecolor{greenF}{HTML}{2CBB5C}
\definecolor{cyan}{HTML}{86A6D5}
\definecolor{darkred}{HTML}{8B0000}

\def\DTZU {2ex}

\tikzstyle{taskfg} = [
  text=deepblue,
]

\tikzstyle{taskbg} = [
  fill=cyan!50,
]

\tikzstyle{taskline} = [
  draw=deepblue,
]

\tikzstyle{taskstyle} = [
  ultra thick,
  taskfg,
  taskbg,
  taskline
]

\tikzstyle{task} = [
  rectangle,
  minimum width=12mm,
  minimum height=10mm,
]

\tikzstyle{smalltask} = [
  rectangle,
  minimum width=8mm,
  minimum height=6mm,
  taskstyle,
  very thick
]

\tikzstyle{constraint} = [
  ultra thick,
  taskline,
]

\tikzstyle{response} = [
  constraint,
  *-triangle 60
]

\tikzstyle{precedence} = [
  constraint,
  *triangle 60-
]
\tikzstyle{succession} = [
  constraint,
  *triangle 60-*
]

\tikzstyle{respondedexistence} = [
  constraint,
  *-
]

\tikzstyle{coexistence} = [
  constraint,
  *-*
]

\tikzstyle{negationconstraint} = [
  constraint,
  postaction={decorate,decoration={markings,
   mark=at position .5 with {\arrow[xshift=0.15*\DTZU]{Bar[width=1.5*\DTZU]}},
   mark=at position .5 with {\arrow[xshift=-0.15*\DTZU]{Bar[width=1.5*\DTZU]}}
  }}
]

\tikzstyle{notcoexistence} = [
  negationconstraint,
  coexistence,
]

\tikzstyle{negationresponse} = [
  constraint,
  response,
  postaction={decorate,decoration={markings,
   mark=at position .5 with {\arrow[xshift=0.15*\DTZU]{Bar[width=1.5*\DTZU]}},
   mark=at position .5 with {\arrow[xshift=-0.15*\DTZU]{Bar[width=1.5*\DTZU]}}
  }}
]

\tikzstyle{negationsuccession} = [
  constraint,
  succession,
  postaction={decorate,decoration={markings,
   mark=at position .65 with {\arrow[xshift=0.15*\DTZU]{Bar[width=1.5*\DTZU]}},
   mark=at position .65 with {\arrow[xshift=-0.15*\DTZU]{Bar[width=1.5*\DTZU]}}
  }}
]

\tikzstyle{exclusivechoice} = [
  constraint,
  -,
  postaction={decorate,decoration={markings,
   mark=at position .5 with {\arrow[xshift=0.5*\DTZU]{Diamond[width=1*\DTZU]}} 
  }}
]

\tikzstyle{choice} = [
  constraint,
  -,
  postaction={decorate,decoration={markings,
   mark=at position .5 with {\arrow[xshift=0.5*\DTZU]{Diamond[width=1*\DTZU]}},
    mark=at position .5 with {\arrow[xshift=0.28*\DTZU,white,scale=.6]{Diamond[width=1*\DTZU]}} 
  }}
]

\tikzstyle{circ} = [
  solid,
  text=black
]

\tikzstyle{timeline} = [
  -,
  thin,
  ]
  
\tikzstyle{snapshot} = [
  -,
  thin,
  densely dotted,
  ]

\tikzstyle{objnode} = [
  inner sep=2pt,
  circle,
  ultra thick,
]

\tikzstyle{tobj} = [
  objnode,
  taskfg,
  taskbg,
  taskline
]

\tikzstyle{cobj} = [
  objnode,
  classfg,
  classbg,
  classline
]

\tikzstyle{link} = [
  solid,
  -angle 60
]

%% file: sections/1.intro.tex

\section{Introduction}
\label{sec:intro}

Linear Temporal Logic (\LTL) is one of the most widely studied modal logics for time. It is interpreted over infinite
state sequences (or traces). Since its introduction by
Pnueli~\cite{pnueli1977temporal}, \LTL has been extensively employed in a variety of verification and synthesis tasks. 
While verification aims at checking the correctness of an \LTL specification
over a given dynamic system, synthesis uses an \LTL specification to derive
a corresponding correct-by-construction program (in the shape, \eg, of a Mealy or Moore machine, I/O-transducer, or circuit) that realizes the specification.
Extensive research has been conducted on different synthesis
settings, considering in particular closed and \emph{open} (also called \emph{reactive})
systems, starting from the seminal contributions by~Harel, Pnueli and Rosner \cite{HarelP84,PnueliR89}.
In the reactive setting, the system (referred to as Controller) interacts with  Environment, which,
in turn, can affect the behavior of Controller. 
Reactive synthesis is hence modeled as a two-player game between
Controller, whose aim is to satisfy the formula, and Environment, who tries
to violate it. The objective of the
synthesis task is then to synthesize a program for Controller indicating which actions the Controller should take to guarantee the satisfaction of the \LTL specification of interest, no matter what are the actions taken by Environment.
This problem was originally studied
in~\cite{church1962logic} and solved in~\cite{buchi1990solving}, and for
\LTL specifications in particular it was shown to be
\EXPTIME[2]-complete~\cite{pnueli1989synthesis,rosner1992modular}. 
The high theoretical complexity and practical infeasibility of the original
approach, based on the reduction to solving parity games on deterministic
parity automata, inspired the scientific community to search for practically
interesting fragments of \LTL for which the synthesis problem is computationally more amenable.
One of the most known examples in this direction is that of Generalized Reactivity(1) formulas~\cite{bloem2012synthesis}, for which the synthesis problem can be solved in cubic time. 

In a variety of application domains, the dynamics of the system consist of unbounded, yet finite, traces \cite{DeGiacomoDM14}. This has led to the introduction of \LTLf \cite{DeGiacomoV13}, \ie \LTL interpreted over \emph{finite} traces. This logic has been extensively studied within AI, formal methods, and business process management~\cite{DeGiacomoV15,PesicA06,DegEtAl,FiondaG18,AMO:IJCAI19,DeGiacomoIFP19,MaMP20,DDMM22}.
In particular, extensive progress has been recently made in \LTLf-based
synthesis~\cite{ZhuTLPV17,CamachoBMM18,ZhuPV19,BansalLTV20,ZhuGPV20,GiacomoSVZ20,XiaoL0SPV21,GiacomoFLVX022},
where the synthesised program always terminates (differently from the infinite-trace case). While dealing with the finite-trace semantics makes the problem much more amenable to algorithmic optimization, it does not change the theoretical complexity of the synthesis problem: \LTLf synthesis is in fact 
\EXPTIME[2]-complete~\cite{DeGiacomoV15}.
The classical algorithm for \LTLf synthesis requires to:
\begin{enumerate*}[label=(\roman*)]
  \item encode the formula into a nondeterministic automaton over finite
    words (\NFA);
  \item turn the \NFA into a deterministic automaton (\DFA), \eg
    through the subset construction algorithm;
  \item solving a \emph{reachability game}, in which the objective of
    Controller is to force the game to reach a final state of the
    automaton.
\end{enumerate*}
Each one of the first two steps requires, in the worst case, an exponential amount of
time in the size of its input.

With the goal of finding \LTLf fragments that at once enjoy a better complexity for synthesis and are meaningful and used in
practice, in this paper we resort on the usage of \LTLf in business process management (BPM). There, \LTLf is employed to define the semantics of one of the most well-studied declarative process modelling languages, namely \DECLARE~\cite{PesicA06,MontaliPACMS10,CiccioM22}. Interestingly, \DECLARE is a pattern-based language: constraints are defined based on a pre-defined set of of unary or binary templates, each of
which is defined as an \LTLf formula. A \DECLARE specification consists of a set of
(conjunctively related) patterns. A flourishing line of research focuses on a variety of reasoning and analysis tasks for \DECLARE, ranging from discovery of \DECLARE specifications from data representing past executions of the process to offline and run-time verification of \DECLARE specifications
and extensions with data~\cite{FiondaG18,CiccioM22,DDMM22}. Notably, \DECLARE patterns are derived from a catalogue of relevant temporal properties in software engineering \cite{DwAC99}, and are closely related to temporal patterns used in planning under trajectory constraints \cite{BacchusK00,GereviniHLSD09}, which makes the study of \DECLARE interesting  beyond BPM. 

Reactive synthesis in this setting comes as a natural problem that, surprisingly, has never been studied in the literature. In the vast majority of cases, the execution of process indeed comes with at least two parties: the organization (and its internal resources) responsible for enacting the process, and one or more external stakeholders taking uncontrollable, constrained decisions on how to progress with the execution. This  calls directly for a form of assume-guarantee synthesis problem for \DECLARE where, given a declarative specification (called \emph{assumption}) regulating how the external activities can be executed, and a declarative specification (called \emph{guarantee}) constraining how internal activities can be executed, one wants to obtain a program (called \emph{orchestration}) indicating to the organization how to ensure that the guarantee is respected under the hypothesis that the external stakeholders behave by respecting the assumption. 

\emph{In this paper, we formalize and study the reactive synthesis problem for
declarative, \DECLARE process specifications}, providing a threefold contribution.
First, we formalize the problem of reactive synthesis of \DECLARE, and illustrate the importance of this problem for the BPM and formal methods communities using a meaningful example. We also give a na\"ive, doubly exponential time algorithm, that reduces the problem to \LTLf synthesis.

Second, we show how to improve the na\"ive algorithm and obtain a singly
exponential time algorithm for \DECLARE synthesis. This refined algorithm is based
on the observation that, starting from pure past temporal formulas, it is
possible to build language-equivalent deterministic finite automata of
singly exponential size~\cite{ChandraKS81,de2021pure}.  In particular, we
introduce for the first time a systematic encoding of all \DECLARE patterns into
linear-size pure-past formulas of \LTLf (we call this procedure
``pastification'', similarly
to~\cite{cimatti2021extended,maler2007synthesizing,maler2005real}).  As
a by-product, our results reveal the following fundamental property: \DECLARE is a fragment of \LTLf with a polynomial pastification
algorithm, a result that, as we show, cannot be achieved for full \LTLf.

Third, we show a novel translation from pure-past temporal formulas to
\emph{symbolic and deterministic} automata, which we use for obtaining
a \emph{symbolic} version of the previous algorithm. This new version represents
\DFAs with boolean formulas and symbolically encodes
reachability games, and paves the way towards a substantial improvement
in performance thanks to the well-known practical benefits of the symbolic
approach~\cite{burch1992symbolic,mcmillan1993symbolic}.
 
The rest of the paper is organized as follows. In \cref{sec:back}, we
review the necessary background: \LTLf
with past and future modalities and its synthesis problem, \DECLARE, finite-state explicit and symbolic automata. 
In \cref{sec:declare:synthesis} we define the reactive synthesis
problem for \DECLARE (together with an example), and show a na\"ive
algorithm for solving it. 
\cref{sec:explicit} presents the singly exponential time algorithm for
\DECLARE synthesis, based on the pastification of \DECLARE formulas.
 \cref{sec:symbolic} shows the symbolic version of the previous
algorithm. Conclusions and future directions follow.
Full proofs for all the Lemmas and Theorems presented
in this paper can be found in the Appendix.


%% file: sections/2.background.tex

\section{Background}
\label{sec:back}
In this section, we give the background necessary for the rest of the
paper.

\subsection{Linear Temporal Logic over finite traces}
\label{sub:back:ltlf}

Given a set $\Sigma$ of proposition letters, a formula $\phi$ of \LTLf is
defined as follows~\cite{DeGiacomoV13}:
\begin{align}
\phi
\coloneqq 
p
& \mid \neg p \mid \phi \lor \phi \mid \phi \land \phi & \text{Boolean connectives} \\
& \mid \ltl{X} \phi \mid \ltl{wX} \phi \mid \phi \ltl{U} \phi \mid \phi \ltl{R} \phi & \text{future modalities} \\
& \mid \ltl{Y} \phi \mid \ltl{wY} \phi \mid \phi \ltl{S} \phi \mid \phi \ltl{T} \phi & \text{past modalities}
\end{align}
where $p \in \Sigma$. Note that the definition of the syntax of \LTLf
generates formulas in \emph{negated normal form} (NNF), that is with
negations appearing only in front of proposition letters.
The future temporal operators $\ltl{X}$, $\ltl{wX}$, $\ltl{U}$, and
$\ltl{R}$ are called \emph{tomorrow}, \emph{weak tomorrow}, \emph{until},
and \emph{release}, respectively. The past temporal operators $\ltl{Y}$,
$\ltl{wY}$, $\ltl{S}$, and $\ltl{T}$ are called \emph{yesterday},
\emph{weak yesterday}, \emph{since}, and \emph{triggers}, respectively.
We use the standard shortcuts for $\top \coloneqq p \lor \lnot p$, $\bot
\coloneqq p \land \lnot p$ (for some $p \in \Sigma$) and for other temporal
operators: $\ltl{F\phi \coloneqq \top U \phi}$ (called \emph{eventually}),
$\ltl{G\phi \coloneqq \false R \phi}$ (called \emph{globally}),
$\ltl{\phi_1 W \phi_2 \coloneqq \phi_1 U \phi_2 \lor G \phi_1}$ (called
\emph{weak until}), $\ltl{O\phi \coloneqq \top S \phi}$ (called
\emph{once}), and $\ltl{H\phi \coloneqq \false T \phi}$ (called
\emph{historically}).

We denote with \LTLfFP the \emph{pure past} fragment of \LTLf, \ie the
fragment of \LTLf devoid of future temporal operators.  With some abuse of
notation, we denote with \LTLf and \LTLfFP also the set of formulas in \LTL
and \LTLfFP, respectively. For any formula $\phi$, we denote with $|\phi|$
the \emph{size} of $\phi$, \ie its number of symbols.

Formulas of \LTLf over the alphabet $\Sigma$ are interpreted over
\emph{finite traces} (or state sequences, or words), \ie sequences in the
set $(2^\Sigma)^{+}$.  In the following, we will write \emph{general finite
traces semantics} to denote the interpretation under this structures.  
Let $\sigma = \langle \sigma_{0}, \ldots, \sigma_{n-1} \rangle \in
(2^{\Sigma})^{+}$ be a finite trace.  We define the \emph{length} of
$\sigma$ as $|\sigma| = n$.
With $\sigma_{[i,j]}$ (for some $0\le i\le j<|\sigma|$) we denote the
subinterval $\seq{\sigma_i,\dots,\sigma_j}$ of $\sigma$.
The \emph{satisfaction} of an \LTLf formula $\phi$ by $\sigma$ at
time $0 \leq i < |\sigma|$, denoted by $\sigma, i \models \phi$, is defined
as follows:
\begin{itemize}
  \item $\sigma,i \models p$ iff $p\in\state_i$;
  \item $\sigma,i \models \ltl{\neg p}$ iff $p\not\in\state_i$;
  \item $\sigma,i \models \ltl{\phi_1 || \phi_2}$  iff
          $\sigma,i \models \phi_1$ or $\sigma,i \models \phi_2$;
  \item $\sigma,i \models \ltl{\phi_1 && \phi_2}$ iff
          $\sigma,i \models \phi_1$ and $\sigma,i \models \phi_2$;
  \item $\sigma,i \models \ltl{X\phi}$     iff 
          $i+1<|\sigma|$ and  $\sigma,i+1\models \phi$;
  \item $\sigma,i \models \ltl{wX\phi}$     iff 
          either $i+1=|\sigma|$ or $\sigma,i+1\models \phi$;
  \item $\sigma,i \models \ltl{Y\phi}$    iff
          $i > 0$ and $\sigma,i-1\models \phi$;
  \item $\sigma,i \models \ltl{wY\phi}$    iff
          either $i = 0$ or $\sigma,i-1\models \phi$;
  \item $\sigma,i \models \ltl{\phi_1 U \phi_2}$  iff
          there exists $i\le j<|\sigma|$ such that $\sigma,j\models\phi_2$,
          and $\sigma,k\models\phi_1$ for all $k$, with $i \le k < j$;
  \item $\sigma,i \models \ltl{\phi_1 S \phi_2}$    iff
          there exists $j\le i$ such that $\sigma,j\models\phi_2$,
          and $\sigma,k\models\phi_1$ for all $k$, with $j < k \le i$;
  \item $\sigma,i \models \ltl{\phi_1 R \phi_2}$  iff either
    $\sigma,j\models\phi_2$ for all $i\le j < |\sigma|$, or there exists
    $i \leq k < | \sigma |$ such that $\sigma,k\models\phi_1$ and $\sigma,j\models\phi_2$
    for all $i\le j \le k$;
  \item $\sigma,i \models \ltl{\phi_1 T \phi_2}$  iff either
    $\sigma,j\models\phi_2$ for all $0\le j \leq i$, or there exists $k \le
    i$ such that $\sigma,k\models\phi_1$ and $\sigma,j\models\phi_2$ for
    all $i\ge j \ge k$.
\end{itemize}
We say that $\sigma$ is a \emph{model} of $\phi$ (written as $\sigma
\models \phi$) iff $\sigma,0 \models \phi$.  The \emph{language} (of finite
words) of $\phi$, denoted by $\lang(\phi)$, is the set of traces
$\sigma\in(2^\Sigma)^+$ such that $\sigma\models\phi$.  We say that two
formulas $\phi, \psi \in \LTLf$ are \emph{equivalent} iff
$\lang(\phi)=\lang(\psi)$. 

If $\phi$ belongs to \LTLfFP (\ie pure past fragment of \LTLf), then we
interpret $\phi$ at the last time point of the trace, \ie we say
that $\sigma \in (2^\Sigma)^+$ is a model of $\phi$ if and only if
$\sigma,|\sigma|-1 \models \phi$.

\mypar{The realizability and reactive synthesis problems of \LTLf}
\label{sub:sub:ltlf:problems}
\emph{Realizability} aims at establishing whether, given an \LTLf  formula
$\phi$ over two sets $\Uset$ and $\Cset$ of respectively uncontrollable and controllable
variables, there exists a strategy which, no matter of
the value of the uncontrollable variables, chooses the value of the
controllable variables so that the formula is satisfied.
\emph{Reactive synthesis} refers to the task of synthesizing such
a strategy in the case the formula is realizable.
Realizability is usually modeled as a two-player game between Environment,
who tries to violate the specification, and Controller, who tries to
fulfill it.

\begin{definition}[Strategy]
\label{def:strategy}
  Let $\Sigma=\Cset\cup\Uset$ be a set of variables partitioned into
  \emph{controllable} $\Cset$ and \emph{uncontrollable} $\Uset$  ones.
  A \emph{strategy} for \emph{Controller} is a function $s:(2^\Uset)^+ \to
  2^\Cset$ that, for any finite sequence
  $\Uncontr=\seq{\Uncontr_0,\ldots,\Uncontr_n}$ of choices by
  \emph{Environment}, determines the choice $\Contr_n=s(\Uncontr)$ of
  \emph{Controller}.
\end{definition}

Let $s:(2^\Uset)^+ \to 2^\Cset$ be a strategy and let
$\Uncontr=\seq{\Uncontr_0,\Uncontr_1,\ldots}$ $\in(2^\Uset)^\omega$ be an
infinite sequence of choices by Environment. We denote by
$\res(s,\Uncontr)= \seq{\Uncontr_0\cup s(\seq{\Uncontr_0}), \Uncontr_1 \cup
s(\seq{\Uncontr_0, \Uncontr_1}), \ldots}$ the trace resulting from
reacting to $\Uncontr$ according to $s$.
The realizability and reactive synthesis problems from \LTLf formulas are
defined as follows.

\begin{definition}[Realizability and Reactive Synthesis for \LTLf]
\label{def:realizability}
  Let $\phi$ be a \LTLf formula over the alphabet $\Sigma = \Cset \cup
  \Uset$, with $\Cset\cap\Uset=\emptyset$. We say that $\phi$ is
  \emph{realizable} (over finite words) if and only if there exists
  a strategy $s : (2^\Uset)^{+} \to 2^\Cset$ such that, for any infinite
  sequence $\Uncontr=\seq{\Uncontr_0, \Uncontr_1, \dots}$ in
  $(2^\Uset)^\omega$, it holds that there exists $k\in\N$ for which
  $\res(s,\Uncontr)_{[0,k]}$ is a model of $\phi$.
  Whenever $\phi$ is realizable, \emph{reactive synthesis} is the
  problem of computing such a strategy $g$~\footnotemark.
\end{definition}

\footnotetext{%
  The strategy $g$ can be concretely represented in many forms, \eg as
  a Mealy or Moore machine~\cite{Vardi95}.
}

Determinism plays a crucial role in realizability. In fact, once 
a \DFA recognizing the language of the formula is obtained, such
automaton can be used as an arena for a game between the Environment and
the Controller players. In the case of automata over finite words,
Controller wants to force the game to visit a final state of the automaton. This type of game is called \emph{reachability game}~\cite{de2007concurrent}
and can be solved through a fixpoint computation in linear time in the size of the arena. We refer to \cite{de2007concurrent,jacobs2017first}
for more details.

The classical algorithm for solving the realizability of an \LTLf formula $\phi$
works as follows~\cite{DeGiacomoV15}:
\begin{enumerate*}[label=(\roman*)]
  \item it builds an \NFA for $\phi$ (\eg using the algorithm
    in~\cite{vardi1994reasoning,vardi1986automata});
  \item it determinizes it (\eg using the subset construction algorithm)
    obtaining a \DFA $\autom$;
  \item it plays a reachability game over $\autom$ to establish whether
    Controller can force the game to reach a final state of the automaton.
\end{enumerate*}
If this is the case, then $\phi$ is realizable, otherwise it is
unrealizable.
The realizability problem for \LTLf is \EXPTIME[2]-complete.

\subsection{The \DECLARE Language}
\label{sub:back:declare}

\DECLARE is a framework and a language for the declarative specification of
processes. We refer to~\cite{montali2010specification} for a thorough
treatment of declarative processes. 

A \DECLARE model (sometimes also called \DECLARE specification) consists of a set of \emph{patterns} used for constraining
the allowed execution traces of the process. Each pattern is defined over
a set of \emph{actions}, \ie atomic tasks representing units of work in the
process. \DECLARE assumes that exactly one action can be executed at each
time point, and that each execution eventually terminates.  For these
reasons, the semantics of \DECLARE is given by means of the \LTLf logic
interpreted over so-called \emph{simple finite traces}~\cite{DegEtAl,fionda2019control}.

A simple finite trace $\sigma$ is a finite trace
$\seq{\sigma_0,\dots,\sigma_n}$ in $\Sigma^+$, \ie such that $|\sigma_i|=1$
for any $0\le i < |\sigma|$. This captures that, at each time point, exactly one proposition
letter must hold, which in the context of \DECLARE indicates which action is executed. In the following, we write \emph{simple finite
traces semantics} to denote the interpretation assuming simple finite
traces.

In \cref{tab:constraints}, for each \DECLARE pattern (first column of the
table) on a set $\mathcal{A}$ of actions, we report the corresponding
formalization with \LTLf formulas over the alphabet
$\Sigma\coloneqq\mathcal{A}$ (second column of the table) assuming
\emph{general} finite trace semantics.  This comes with no loss of
generality, since one can always pass from the general to the simple finite traces semantics by adding the following \LTLf constraint (by analogy with~\cite{FiondaG18}):
$\mathtt{simple}(\Sigma) \coloneqq \ltl{G(\bigvee_{p\in\Sigma} p \land
\bigwedge_{p \neq q \in \Sigma}!(p & q))}$.

From now on, with some abuse of notation, with ``\DECLARE pattern'' we
refer to its corresponding formalization in \LTLf. Similarly, with
``\DECLARE model'' we refer to a conjunction of \DECLARE patterns.

\input{figures/declare-patterns.tex}

\subsection{Automata}
\label{sub:back:automata}

Automata are among the first computational models that have been
introduced, and have a strong connection with temporal logics and formal
verification (see, for instance, \cite{vardi1994reasoning}).
In this paper, we focus our attention on automata reading \emph{finite
words}. Given an alphabet $\Sigma$, a finite word $\sigma$ is a member of
$\Sigma^*$. From now on, we write ``word'' to denote a finite word.  We
use two representations of automata, the explicit-state one (which is
the classical) and the symbolic one. 

\mypar{Explicit-state automata}
We recall automata over finite words.

\begin{definition}[\NFAs and \DFAs]
\label{def:nfa}
  A \emph{nondeterministic finite automaton} (\NFA, for short) $\autom$ is
  a tuple $\tuple{\Sigma,Q,I,\delta,F}$ such that:
  \begin{enumerate*}[label=(\roman*)]
    \item $\Sigma$ is a finite alphabet;
    \item $Q$ is a set of states;
    \item $I\subseteq Q$ is the set of initial states;
    \item $\delta : Q \times \Sigma \to 2^Q$ is the transition relation;
    \item $F\subseteq Q$ is the set of final states.
  \end{enumerate*}
  A \emph{deterministic finite automaton} (\DFA, for short) is an \NFA such
  that $|I|=1$ (\ie there exists exactly one initial state) and $\delta$ is
  a function (\ie $\delta : Q \times \Sigma \to Q$).
\end{definition}

Given an \NFA $\autom$ with set of states $Q$, we denote with $|\autom|$
the number of its states (\ie $|Q|$). 
Given an \NFA $\autom=\tuple{\Sigma,Q,I,\delta,F}$ and a word
$\sigma=\seq{\sigma_0,\dots,\sigma_n}\in\Sigma^*$, a \emph{run $\pi$
induced by $\sigma$ in $\autom$} is a (finite) sequence of states
$\seq{q_0,\dots,q_{n+1}}\in Q^*$ such that $q_0 \in I$ and $q_{i+1} \in
\delta(q_i,\sigma_i)$, for any $i\ge 0$. In a \DFA, for each word $\sigma$,
there is exactly one run induced by $\sigma$.  
A run $\pi=\seq{q_0,\dots,q_{n+1}}$ is \emph{accepting} iff $q_{n+1} \in
F$.  
The language of $\autom$, denoted as $\lang(\autom)$, is the set of words
$\sigma$ such that there exists at least one accepting run induced by
$\sigma$ in $\autom$.

The set of languages recognizable by \NFAs coincides with the set of
languages recognizable by \DFAs. This is because \NFAs can be determinized
(\ie turned into language-equivalent \DFAs) with the subset/powerset
construction algorithm~\cite{hopcroft2001introduction}. In the worst case,
this algorithm produces a \DFA of exponential size with respect to the
starting \NFA.
%
%
Each language expressible in \LTLf can be recognized by an \NFA, whose size
is exponential in the size of the formula, in the worst
case~\cite{DeGiacomoV13}.  Together with the subset construction algorithm,
one can obtain a translation from \LTLf formulas to language-equivalent
\DFAs of doubly exponential size~\footnotemark.
\footnotetext{%
  Note that the converse does not hold: \DFAs recognize
  a strictly larger set of languages. In particular, \LTLf corresponds to
  the counter-free fragment of \DFAs~\cite{mcnaughton1971counter}.
}
Interestingly, although having the same expressive power of
\LTLf~\cite{lichtenstein1985glory,zuck1986past,thomas1988safety}, the
\LTLfFP logic admits a construction of equivalent \DFAs of singly
exponential size.  The crucial observation is that, since ``past already
happened'', pure past formulas can be turned into automata without the need
of introducing nondeterminism.

\begin{proposition}[\cite{de2021pure}]
\label{prop:ltlfp:dfa}
  For any formula $\phi\in\LTLfFP$ of size $n=|\phi|$, there exists a \DFA
  $\autom$ such that $\lang(\phi)=\lang(\autom)$ and
  $|\autom|\in2^{\mathcal{O}(n)}$.
\end{proposition}

\mypar{Symbolic automata}
The drawback of the explicit-state representation of automata is the size
that it requires. In some application, \eg model checking or reactive
synthesis, it can be that the automaton has a set of states and a set of
transitions that are prohibitively large to be represented in memory.  The
symbolic representation has the objective to overcome this issue by
representing automata by means of Boolean
formulas~\cite{mcmillan1993symbolic}: states and transitions are never
explicitly represented in memory but, on the contrary, their are
symbolically represented (hence the name) as models of some Boolean
formulas. On the average case, this representation can be exponentially
more succinct than the classical one.  Problems like model checking and
synthesis can thus manipulate automata by manipulating Boolean formulas. 
We take the definitions of this section
from~\cite{phdgeatti,cimatti2021extended,cimatti2021fairness}.  We first
give the definition of a \emph{symbolic \NFA}.

\begin{definition}[Symbolic \NFAs]
\label{def:automaton}
  A \emph{symbolic \NFA} over the alphabet $\Sigma$ is a tuple $\autom = (X
  \cup \Sigma, I(X), T(X,\Sigma,X'), F(X,\Sigma))$, where
  \begin{enumerate*}[label=(\roman*)]
    \item $X$ is a set of \emph{state variables},
    \item $I(X)$ and $T(X,\Sigma, X^\prime)$, with $X'=\set{x' \mid x\in X}$, are
      Boolean formulas which define the set of initial states and the transition
      relation, respectively, and  
    \item $F(X,\Sigma)$ is a Boolean formula over the variables in
      $X\cup\Sigma$ which defines the set of final states.
  \end{enumerate*}
\end{definition}

Given a symbolic \NFA $\autom= (X\cup\Sigma, I(X), T(X,\Sigma,X'),
F(X,\Sigma))$, we denote with $|\autom|$ the sum of the number of
symbols in $I$, $T$ and $F$ (\ie $|I|+|T|+|F|$).  Symbolic \NFAs
can be refined into symbolic \DFAs like for explicit-state automata. In the
symbolic representation, a \DFA is such that its formula $I(X)$ has exactly
one satisfying assignment and, for each state variable $x$, the value of
$x$ at the next state is uniquely determined by the value of the state and
input variables in the current state.

\begin{definition}[Symbolic \DFAs]
\label{def:sdfa}
  A \emph{symbolic \DFA} $\autom = (V, I(X), T(X,\Sigma,X^\prime), \\
  F(X,\Sigma))$ is a symbolic \NFA such that
  \begin{enumerate*}[label=(\roman*)]
  \item the formula $I(X)$ has exactly one satisfying assignment, and
  \item the transition relation is of the form:
  \end{enumerate*}
        \[T(X,\Sigma,X^\prime) \coloneqq 
          \bigwedge_{x \in X}(x^\prime \iff \beta_x(V))\]
  where $\beta_x(V)$ is a Boolean formula over $V (= X \cup \Sigma)$, for each $x
  \in X$.
\end{definition}

Given a symbolic \NFA $\autom = (V, I(X), T(X,\Sigma,X^\prime),
F(X,\Sigma))$, a run $\tau = \seq{\tau_0,\dots,\tau_n}$ (for some
$n\in\N$) is a finite sequence of pairs $\tau_i \coloneqq (X_i,\Sigma_i)
\subseteq 2^X \times 2^\Sigma$ (representing the state variables and the
input variables that are supposed to hold at time point $i$) that satisfies
the following two conditions:
\begin{enumerate*}[label=(\roman*)]
  \item $\tau_0 \models I(X)$;
  \item $\tau_i,\tau_{i+1}\models T(X,\Sigma,X^\prime)$, for each $0\le
    i \le n$, when $\tau_i$ is used for interpreting the variables in $X$
    and $\Sigma$ and $\tau_{i+1}$ is used for interpreting the variables in
    $X^\prime$.
\end{enumerate*}

A run $\tau=\seq{(X_0,\Sigma_0),\dots,(X_{n},\Sigma_{n})}$ is induced
by the word $\seq{\sigma_0,\dots,\sigma_n}\in\Sigma^*$ (for some
$n\in\N$) iff $\sigma_i = \Sigma_i$ for all $0\le i\le n$.
A run $\tau$ is \emph{accepting} iff $\tau_n \models F(X,\Sigma)$.
A word $\sigma$ is \emph{accepted} by $\autom$ iff there
exists an accepting run induced by $\sigma$ in $\autom$.  The
\emph{language} of $\autom$, denoted by $\lang(\autom)$, is the set of all
and only the infinite words accepted by $\autom$.

The symbolic representation does not change the expressive power of \NFAs
or \DFAs. In fact, given an \NFA $\autom$ one can easily build the formulas
for its set of initial states, transitions, and final states and obtaining
a language-equivalent symbolic \NFA. Similarly, given a symbolic \NFA, one
can take all models of the formulas $I(X)$, $T(X,\Sigma,X^\prime)$, and
$F(X,\Sigma)$ and use them as states and transitions for
a language-equivalent \NFA.  As mentioned above, what changes between the
two represetations is that symbolic automata can be exponentially more
succinct than explicit-state automata, and this is particularly useful for
applications.

%% file: figures/declare-patterns.tex

\renewcommand{\arraystretch}{1.5}

\renewcommand{\cellalign}{tl}
\renewcommand{\theadalign}{tl}

\begin{table}[h!]
\centering
\begin{tabular}{ p{3.8cm}p{3.4cm}p{4.8cm} } 
 \hline
  \textbf{Pattern} & \textbf{\LTLf formalization} & \textbf{Pastification} ($\pastify(\cdot)$) \\ \hline\hline
  \texttt{existence}($p$)                                                 & 
    $\ltl{F(p)}$                                                          & 
    \cellcolor{lightgray}$\ltl{O(p)}$                                                          \\ \hline
  \texttt{absence2}($p$)                                                  & 
    $\ltl{! F(p & X F(p))}$                                               & 
    \cellcolor{lightgray}$\ltl{H(p -> wY H (! p))}$                                            \\ \hline
  \texttt{choice}($p,q$)                                                  & 
    $\ltl{F(p) | F(q)}$                                                   & 
    \cellcolor{lightgray}$\ltl{O(p | q)}$                                                      \\ \hline
  \texttt{exc-choice}($p,q$)                                              & 
    \makecell{$\ltl{(F(p) | F(q)) & {}}$ \\ $\ltl{!(F(p) & F(q))}$}       & 
    \cellcolor{lightgray}\makecell{$\ltl{O(p | q) & {}}$ \\ $\ltl{(H(!p) | H(!q))}$}           \\ \hline
  \texttt{resp-existence}($p,q$)                                          & 
    $\ltl{F(p) -> F(q)}$                                                  & 
    \cellcolor{lightgray}$\ltl{H(!p) | O(q)}$                                                  \\ \hline
  \texttt{coexistence}($p,q$)                                             & 
    $\ltl{F(p) \iff F(q)}$                                                & 
    \cellcolor{lightgray}\makecell{$\ltl{(H(!p) | O(q)) & {}}$ \\ $\ltl{(H(!q) | O(p))}$}          \\ \hline
  \texttt{response}($p,q$)                                                & 
    $\ltl{G(p -> F(q))}$                                                  & 
    \cellcolor{lightgray}$\ltl{q T (!p | q)}$                                                  \\ \hline
  \texttt{precedence}($p,q$)                                              & 
    $\ltl{(! q) W (p)}$                                                   & 
    \cellcolor{lightgray}$\ltl{H(q -> O(p))}$                                                  \\ \hline
  \texttt{succession}($p,q$)                                              & 
    \makecell{$\ltl{G(p -> F(q)) & {}}$ \\ $\ltl{(! q) W (p)}$}           & 
    \cellcolor{lightgray}\makecell{$\ltl{p T (!p | q) & {}}$ \\ $\ltl{H(q -> O(p))}$}          \\ \hline
  \texttt{alt-response}($p,q$)                                            & 
    $\ltl{G(p -> X((! p) U q))}$                                          & 
    \cellcolor{lightgray}$\ltl{(p | q) T (!p) & H(q -> wY(q T ((p & !q) -> Z(q T !p))))}$      \\ \hline
  \texttt{alt-precedence}($p,q$)                                                                              &
    \makecell{$\ltl{((! q) W p) & {}}$ \\ $\ltl{G(q -> wX((! q) W p))}$}                                      & 
    \cellcolor{lightgray}\makecell{$\ltl{H(q -> O(p)) & {}}$ \\ $\ltl{H((q & !p) -> {}}$ \\ $\ltl{wY (p T (q -> (p T (! p)))))}$}       \\ \hline
  \texttt{alt-succession}($p,q$)                                                                              & 
    \makecell{$\ltl{G(p -> X((! p) U q)) & {}}$ \\ $\ltl{((! q) W p) & {}}$ \\ $\ltl{G(q -> wX((! q) W p))}$}  & 
    \cellcolor{lightgray}\makecell{$\pastify$(\texttt{alt-response}($p,q$)) $\land {}$ \\ $\pastify$(\texttt{alt-precedence}($p,q$))}  \\ \hline
  \texttt{chain-response}($p,q$)                                          & 
    $\ltl{G(p -> X(q))}$                                                  & 
    \cellcolor{lightgray}$\ltl{! p & H(Y(p) -> q)}$                                            \\ \hline
  \texttt{chain-precedence}($p,q$)                                        & 
    $\ltl{G(X(q) -> p)}$                                                  & 
    \cellcolor{lightgray}$\ltl{H(q -> wY p)}$                                                  \\ \hline
  \texttt{chain-succession}($p,q$)                                                                                &
    $\ltl{G(p \iff X(q))}$                                                                                        &
    \cellcolor{lightgray}\makecell{$\pastify$(\texttt{chain-response}($p,q$)) $\land {}$ \\ $\pastify$(\texttt{chain-precedence}($p,q$))}  \\ \hline
  \texttt{not-coexistence}($p,q$)                                         & 
    $\ltl{!(F(p) & F(q))}$                                                & 
    \cellcolor{lightgray}$\ltl{H(!p) | H(!q)}$                                                 \\ \hline
  \texttt{neg-succession}($p,q$)                                          & 
    $\ltl{G(p -> ! F(q))}$                                                & 
    \cellcolor{lightgray}$\ltl{H(!p) | (!q)S(p & !q & wY H(!p))}$                              \\ \hline
  \makecell{\texttt{neg-chain-} \\ \texttt{succession}($p,q)$}                      &
    \makecell{$\ltl{G(p -> wX(! q))} \land {}$ \\ $\ltl{G(q -> wX(! p))}$}          &
    \cellcolor{lightgray}\makecell{$\ltl{H(Y(p) -> ! q)} \land {}$ \\ $\ltl{H(Y(q) -> ! p)}$}            \\ \hline
 \hline
\end{tabular}
\caption{Name of \DECLARE patterns along with their formalization in \LTLf
and their pastification. The grey cells are one of the contributions of the
paper.}
\label{tab:constraints}
\end{table}

%% file: sections/3.declaresynthesis.tex

\section{Reactive Synthesis of \DECLARE}
\label{sec:declare:synthesis}

We now define realizability and reactive synthesis
 for \DECLARE, describe an illustrative example, and provide
a na\"ive algorithm for solving these problems. 

\subsection{Realizability over Simple Traces}
We start from the realizability problem for \LTLf interpreted over simple
finite traces. This significantly differs from the one of \LTLf over general
finite traces (see \cref{def:realizability}) in two main respects.  Firstly,
in the case of simple traces, we impose that both players can play only
one proposition letter from their set.  Secondly, we impose a strict
alternation: Environment starts to play at the first time
point and whenever Environment plays at time point $i$, Controller plays at
time point $i+1$, if the play has not finished before.

We begin by defining the basic building blocks.  A \emph{simple strategy}
for Controller is a function $s:(\Uset)^+ \to \Cset$ that, for every finite
sequence of elements in $\Uset$, determines an element in the set $\Cset$.
Let $s:(\Uset)^+ \to \Cset$ be a simple strategy and let
$\Uncontr=\seq{\Uncontr_0,\Uncontr_1,\ldots}$ $\in(\Uset)^\omega$ be an
infinite \emph{simple} trace of choices by Environment. We denote by
$\simpleres(s,\Uncontr)$ the state sequence $\seq{\Uncontr_0,
s(\seq{\Uncontr_0}), \Uncontr_1, s(\seq{\Uncontr_0, \Uncontr_1}), \ldots}$ resulting from the alternation between the choices of Environment and the
corresponding choices of $s(\cdot)$. Crucially, for any $k\ge 0$,
$\simpleres(s,\Uncontr)_{[0,k]}$ is a \emph{simple} finite trace.
We then define realizability in this setting as follows.

\begin{definition}[Realizability of \LTLf over simple finite traces]
\label{def:ltlf:simple:real}
  Let $\phi$ be a \LTLf formula over the alphabet $\Sigma=\Uset\cup\Cset$
  such that $\Uset\cap\Cset=\emptyset$.  The formula $\phi$ is
  \emph{realizable over simple finite traces} if and only if there exists
  a simple strategy $s : (\Uset)^{+} \to \Cset$ such that, for any infinite
  sequence $\Uncontr=\seq{\Uncontr_0, \Uncontr_1, \dots}$ in
  $(\Uset)^\omega$, there exists $k\in\N$ such that
  $\simpleres(s,\Uncontr)_{[0,k]} \models \phi$.
\end{definition}

The use of a strict alternation between the rounds of the two players appears as the most natural choice when considering \emph{simple finite} traces. On the
one hand, the two players cannot play at the same round, since this would
result in a non-simple trace. On the other hand, giving the possibility to
one of the two players to play for a round of unbounded length would call for additional, technical considerations related to fairness and stability \cite{ZhuGPV20}.
%
We remark that rounds of any bounded length can be simulated with
\cref{def:ltlf:simple:real} by introducing two auxiliary variables that
model a \texttt{no-op} action of the two players, and by the addition of
\LTLf constraints that, whenever the round belongs to one of the two players, forces the other to choose the \texttt{no-op} action. In addition, even in the strict alternating case of \cref{def:ltlf:simple:real}, assigning an unconstrained, \texttt{no-op} action to each of the players is useful to give them the ability to choose of simply release control to the other player.


\subsection{\DECLARE Realizability}

We now turn to realizability for \DECLARE. Differently from the case of standard synthesis, and similarly to the case of agents in AI \cite{ZhuD22}, also in BPM there is background knowledge on how Environment operates. In particular, in BPM the external stakeholders participate to the execution of the process in a constrained way: when it is their turn, they take (arbitrary) decisions on which next action to trigger but only on the subset of all available actions that is made available by the information system supporting the enactment of the process \cite{DRMR18}. 

In this light, \DECLARE calls for an \emph{assume-guarantee} paradigm to synthesis, in the spirit of \cite{CamachoBM18}: Controller guarantees to enforce certain \DECLARE constraints, under the assumption that Environment satisfies other \DECLARE constraints. More specifically, the
realizability problem takes as input two \DECLARE models, one for the
assumptions and the other for the guarantees to fulfill, and aims at finding a strategy for Controller fulfilling all its constraints, regardless on how Environment behaves in the space of possibilities given by the assumption constraints. Environment looses the
realizability game if it violates its assumptions.

\begin{definition}[Realizability and Reactive Synthesis for \DECLARE]
\label{def:declare:real}
  Given two \DECLARE models $\phi_E$ and $\phi_C$ respectively representing the environment and controller specifications, the realizability
  problem of $(\phi_E,\phi_C)$ is the problem of establishing whether
  (the formalization in \LTLf of) $\phi_E \to \phi_C$ is realizable over
  simple finite traces.
  Whenever it is realizable, \emph{reactive synthesis} is the problem of
  computing a simple strategy.
\end{definition}

The use of this \emph{assume-guarantee} approach is particularly important for \DECLARE, in the light of the fact that it does not adopt the full expressiveness of \LTLf, but focuses instead on conjunctively-related patterns of formulas. The following example illustrates this aspect in detail.



\input{sections/example.tex}
\begin{example}
\label{ex:declare}
Consider the (fragment of) an order-to-delivery \DECLARE process depicted
  in~\cref{fig:declare}. The process is enacted by an online shop company,
  by interacting with an external customer interested in ordering material
  from the shop. In this light,
the actions contained in the process are partitioned into those under the responsibility of the online shop (\eg, shipping the order, canceling it or opting for a partial refund to the customer) -- shown in orange, and those under the control of the customer (\eg, opening and closing an order, adding items to it, paying or sending a cancelation request) -- in light blue. 

We then have that customer plays the role of Environment, and the online
  shop that of Controller. Constraints referring to customer actions only
  (in red), or binary constraints having a shop action as source and
  a customer action as target (in green), form the environment
  specification (i.e., the assumption under which the online shop operates). Constraints referring to shop actions only (in blue), or binary constraints having a customer action as source and a shop action as target (in magenta), form the shop specification (i.e., the  constraints the shop is committed to guarantee towards the customer).

More specifically, the process goes as follows. The customer can register their address ($\mathit{regAddr}$), open an order ($\mathit{open}$),  
fill the order with items ($\mathit{add}$), 
close the order ($\mathit{close}$) and pay for it ($\mathit{pay}$), or 
request its cancelation ($\mathit{reqCancel}$).
Whenever an order is open, the customer has to eventually register their address (unless this has been already done). Further address registrations can be performed to update the address data. This is captured by the \texttt{resp-existence}($\mathit{open}$, $\mathit{regAddr}$) constraint. 
Once an order is open, it can be filled with items (\texttt{precedence}($\mathit{open}$, $\mathit{add}$)). Once at least one item is added (i.e., the order is non-empty), the customer can choose to close the order (\texttt{precedence}($\mathit{add}$, $\mathit{close}$)). Closing the order has a twofold effect: on the one hand, no further items can be added (\texttt{neg-succession}($\mathit{close}$, $\mathit{add}$)), on the other hand, it becomes possible to pay the order (\texttt{precedence}($\mathit{close}$, $\mathit{pay}$)) in a single installment (\texttt{abscence2}($\mathit{pay}$)). Finally, a payment can be performed only if the customer has not previously issued a request for order cancelation  (\texttt{neg-succession}($\mathit{reqCancel}$, $\mathit{pay}$)).

On the seller side of the process, the shop can ship the order 
($\mathit{ship}$) and deal with its full cancelation ($\mathit{cancel}$) or partial refund ($\mathit{refund}$). The shop is also equipped with a no-op action ($\mathit{skip}$) to deliberately return control to the customer. 
The portion of process under the responsibility of the shop is
regulated by the following contractual constraints. First of all, full and partial cancelation are mutually exclusive ((\texttt{not-coexistence}($\mathit{cancel}$, $\mathit{refund}$)). A full cancelation can be selected only if the order has not been already shipped (\texttt{neg-succession}($\mathit{ship}$,
$\mathit{cancel}$)), while partial refund is enabled only for shipped orders (\texttt{precedence}($\mathit{cancel}$, $\mathit{refund}$)).



Beside constraints applying only to customer actions or shop actions, there
  are also constraints mutually relating the actions of the two parties.
  These constraint form a sort of binding contract between the two. On the
  other hand, the customer agrees that their address cannot be changed
  anymore once the order is shipped
  (\texttt{neg-succession}($\mathit{ship}$, $\mathit{regAddr}$)). On the
  other hand, the shop guarantees that whenever an order is paid, it will be shipped, and that a shipment only occurs upon a prior payment (\texttt{succession}($\mathit{pay}$, $\mathit{ship}$)). In addition, a shipment can only occur upon a prior execution of at least one address registration by the customer (\texttt{precedence}($\mathit{regAddr}$, $\mathit{ship}$)). Finally, the shop commits that whenever the customer requests to cancel the order, a full cancelation or a partial refund will be performed. This last situation is modelled using a response pattern with a disjunction in the head (\texttt{response}($reqCancel$, $refund\lor cancel$)) -- a feature supported by \DECLARE \cite{PesicA06,montali2010specification}.
\end{example}

%

\subsection{Na\"ive Algorithm for \DECLARE synthesis}

We show a na\"ive algorithm for solving realizability and reactive
synthesis from \DECLARE specifications. The algorithm reduces the problem to
realizability and synthesis of \LTLf over general finite traces semantics.
We first define the following \LTLf formulas, enforcing
\begin{enumerate*}[label=(\roman*)]
  \item the game of the play to be a simple trace;
  \item the Environment player to play at all even time points;
  \item the Controller player to play at all odd time points~\footnotemark.
\end{enumerate*}
\footnotetext{%
  Note the role of the \emph{weak tomorrow} operators in
  $\mathtt{simple}_{\mathtt{Env}}(\Uset)$ and
  $\mathtt{simple}_{\mathtt{Con}}(\Cset)$, which ensure that the game can
  stop the game at any moment.
}
\begin{align}
  \mathtt{simple}_{\mathtt{Env}}(\Uset) &\coloneqq
    \bigvee_{u\in\Uset}u \land \ltl{G}(\bigvee_{u\in\Uset}u \to
    (\bigwedge_{u\not=u'\in\Uset} \lnot(u \land u') \land
    \ltl{wX}(\bigwedge_{u\in\Uset} \lnot u \land \ltl{wX}
    \bigvee_{u\in\Uset} u))) \\
  \mathtt{simple}_{\mathtt{Con}}(\Cset) &\coloneqq
    \bigwedge_{c\in\Cset}\lnot c \land \ltl{G}(\bigwedge_{c\in\Cset}\lnot
    c \to \ltl{wX}(\bigvee_{c\in\Cset} c \land \bigwedge_{c\not=c'\in\Cset}
    \lnot (c \land c') \land \ltl{wX} \bigwedge_{c\in\Cset} \lnot c)))
\end{align}

In the following lemma, we prove that \DECLARE realizability can be reduced
to \LTLf realizability (over general finite traces semantics) by means of
a careful addition of the previous two formulas.

\begin{restatable}{lemma}{lemmasimplegeneralreal}
\label{lemma:simple:general:real}
  Let $\phi_E$ and $\phi_C$ be two \DECLARE models over the set of actions
  $\Sigma = \Uset \cup \Cset$. It holds that $(\phi_E,\phi_C)$ is
  realizable iff the \LTLf formula 
  $\mathtt{simple}_{\mathtt{Con}}(\Cset) \land
  ((\mathtt{simple}_{\mathtt{Env}}(\Uset) \land \phi_E) \to \phi_C)$
  is realizable.
\end{restatable}
\begin{proof}
  We begin proving the left-to-right direction, \ie $(\phi_E,\phi_C)$ is
  realizable implies that the formula
  $\mathtt{simple}_{\mathtt{Con}}(\Cset) \land
  ((\mathtt{simple}_{\mathtt{Env}}(\Uset) \land \phi_E) \to \phi_C)$ is
  realizable.

  Suppose that $(\phi_E,\phi_C)$ is realizable.
  By \cref{def:declare:real}, it holds that there exists a simple strategy
  $s: (\Uset)^+ \to \Cset$ such that, for any
  $\Uncontr=\seq{\Uncontr_0,\Uncontr_1,\dots} \in (\Uset)^\omega$, there
  exists a $k\ge 0$ for which $\simpleres(s,\Uncontr)_{[0,k]} \models
  \phi_E \to \phi_C$.
  We define the strategy $s' : (2^\Uset)^+ \to 2^\Cset$ as the strategy for
  Controller that agrees with $s$ on the odd positions (\ie the ones in
  which Controller moves, under simple trace semantics) and plays the empty
  set on all the other positions (\ie the ones in which Environment moves,
  under simple trace semantics):
  \begin{align}
    s'(\seq{\Uncontr_0,\dots,\Uncontr_j})=
    \begin{cases}
      s(\seq{\Uncontr_0,\dots,\Uncontr_{\frac{j-1}{2}}}) & \mbox{if
      } j \mbox{ is odd} \\
      \emptyset & \mbox{otherwise}
    \end{cases}
  \end{align}
  for any $j\ge 0$. From now on, let
  $\Uncontr'=\seq{\Uncontr'_0,\Uncontr'_1,\dots}$ be an arbitrary sequence
  in $(2^\Uset)^\omega$.  
  We show that $\res(s',\Uncontr')_{[0,k']} \models
  \mathtt{simple}_{\mathtt{Con}}(\Cset) \land
  ((\mathtt{simple}_{\mathtt{Env}}(\Uset) \land \phi_E) \to \phi_C)$ for
  some $k'\ge 0$.

  First of all, we note that, by construction of $s'(\cdot)$ and since
  $s(\cdot)$ is a \emph{simple strategy}, it holds that
  $\res(s',\Uncontr')_{[0,j]} \models
  \mathtt{simple}_{\mathtt{Con}}(\Cset)$, for all $j\ge 0$. Therefore, we
  have to show that $\res(s',\Uncontr')_{[0,k']} \models
  (\mathtt{simple}_{\mathtt{Env}}(\Uset) \land \phi_E) \to \phi_C$ for some
  $k'\ge 0$.
  We divide in cases:
  \begin{itemize}
    \item Suppose that at least one of these conditions holds: 
      \begin{enumerate*}[label=(\roman*)]
        \item $|\Uncontr'_i| \neq 1$ for some $i\ge 0$ such that $i$ is
          even;
        \item $|\Uncontr'_i| \neq 0$ for some $i\ge 0$ such that $i$ is
          odd.
      \end{enumerate*}
      Then, $\res(s',\Uncontr')_{[0,k']} \not\models
      \mathtt{simple}_{\mathtt{Env}}(\Uset)$, where $k'=i$, and thus
      $\res(s',\Uncontr')_{[0,k']} \models
      (\mathtt{simple}_{\mathtt{Env}}(\Uset) \land \phi_E) \to \phi_C$.
    \item Suppose that there exists a $k'\ge 0$ such that
      $\res(s',\Uncontr')_{[0,k']} \not\models \phi_E$. In this case,
      $\res(s',\Uncontr')_{[0,k']} \models
      (\mathtt{simple}_{\mathtt{Env}}(\Uset) \land \phi_E) \to \phi_C$.
    \item Otherwise, for all $j\ge 0$, it holds that 
      \begin{enumerate*}[label=(\roman*)]
        \item $\res(s',\Uncontr')_{[0,j]} \models
          \mathtt{simple}_{\mathtt{Env}}(\Uset)$; and
        \item $\res(s',\Uncontr')_{[0,j]} \models \phi_E$.
      \end{enumerate*}
      It remains to show that $\res(s',\Uncontr')_{[0,k']} \models \phi_C$
      for some $k' \ge 0$. 

      We define $\Uncontr''$ as the sequence in $(\Uset)^\omega$ obtained
      from $\Uncontr'$ by removing all odd positions:
      \begin{align}
        \Uncontr'' \coloneqq
        \seq{\Uncontr'_i \suchthat i \mbox{ is even}}
      \end{align}
      By hypothesis, we know that $\simpleres(s,\Uncontr'')_{[0,k]} \models
      \phi_C$ for some $k\ge 0$. Moreover, since
      $\res(s',\Uncontr')_{[0,j]} \models
      \mathtt{simple}_{\mathtt{Env}}(\Uset) \land
      \mathtt{simple}_{\mathtt{Con}}(\Cset)$ for all $j\ge 0$, it holds
      that $\res(s',\Uncontr') = \simpleres(s,\Uncontr'')$.  
      It follows that $\res(s',\Uncontr')_{[0,k']} \models \phi_C$ for
      $k'=k$.
  \end{itemize}

  We now prove the right-to-left direction. Suppose that the \LTLf formula
  $\mathtt{simple}_{\mathtt{Con}}(\Cset) \land
  ((\mathtt{simple}_{\mathtt{Env}}(\Uset) \land \phi_E) \to \phi_C)$ is
  realizable.  We prove that $(\phi_E,\phi_C)$ is realizable as well.
  By hypothesis, there exists a strategy $s : (2^\Uset) \to 2^\Cset$ such
  that, for any $\Uncontr=\seq{\Uncontr_0,\Uncontr_1,\dots}$, there exists
  a $k\ge 0$ for which $\res(s,\Uncontr)_{[0,k]} \models
  \mathtt{simple}_{\mathtt{Con}}(\Cset) \land
  ((\mathtt{simple}_{\mathtt{Env}}(\Uset) \land \phi_E) \to \phi_C)$.
  We define the strategy $s' : (\Uset)^+ \to \Cset$ as follows:
  \begin{align}
    s'(\seq{\Uncontr_0,\dots,\Uncontr_k}) \coloneqq 
    s(\seq{\Uncontr_0,\dots,\Uncontr_{2\cdot k+1}})
  \end{align}
  Note that, since by hypothesis $s(\cdot)$ is a strategy that realizes
  $\mathtt{simple}_{\mathtt{Con}}(\Cset) \land
  ((\mathtt{simple}_{\mathtt{Env}}(\Uset) \land \phi_E) \to \phi_C)$, in
  particular it holds that $s(\cdot)$ realizes
  $\mathtt{simple}_{\mathtt{Con}}(\Cset)$, and
  $|s(\seq{\Uncontr_0,\dots,\Uncontr_j})| = 1$ for any $j\ge 0$ such that
  $j$ is odd, and thus $s'$ is a \emph{simple strategy}.

  From now on, let $\Uncontr'=\seq{\Uncontr'_0,\Uncontr'_1,\dots}$ be any
  sequence in $(\Uset)^\omega$. We show that there exists a $k'\ge 0$ for
  which $\simpleres(s',\Uncontr')_{[0,k']} \models \phi_E \to \phi_C$.
  We divide in cases:
  \begin{itemize}
    \item Suppose that there exists a $k'\ge 0$ for which
      $\simpleres(s',\Uncontr')_{[0,k']} \not\models \phi_E$; then, of course,
      $\simpleres(s',\Uncontr')_{[0,k']} \models \phi_E \to \phi_C$.
    \item Otherwise, it holds that $\simpleres(s',\Uncontr')_{[0,k']}
      \models \phi_E$ for all $k'\ge 0$.
      We define
      $\Uncontr''$ as the sequence in $(2^\Uset)^\omega$ obtained from
      $\Uncontr'$ by interleaving the empty set between each pair of successive
      indices, that is:
      \begin{align}
        \Uncontr'' \coloneqq \seq {\Uncontr'_0, \emptyset, \Uncontr'_1,
        \emptyset, \Uncontr'_2, \emptyset, \dots}
      \end{align}
      By hypothesis, it holds that $\res(s,\Uncontr'')_{[0,k]} \models
      \mathtt{simple}_{\mathtt{Con}}(\Cset) \land
      ((\mathtt{simple}_{\mathtt{Env}}(\Uset) \land \phi_E) \to \phi_C)$
      for some $k\ge 0$.  
      Moreover, by construction of $\Uncontr''$, it holds that
      $\res(s,\Uncontr'')_{[0,k]} \models
      \mathtt{simple}_{\mathtt{Env}}(\Uset)$.
      Together with the fact that $\res(s,\Uncontr'')_{[0,k]} \models
      \mathtt{simple}_{\mathtt{Con}}(\Cset)$, this implies that
      $\simpleres(s',\Uncontr') = \res(s,\Uncontr'')$.
      We divide again in cases:
      \begin{itemize}
        \item Suppose that $\res(s,\Uncontr'')_{[0,k]} \not\models \phi_E$
          for some $k\ge 0$. This is a contradiction, because this implies
          that $\res(s',\Uncontr')_{[0,k']} \not\models\phi_E$ (for $k'=k$)
          but we have supposed that $\simpleres(s',\Uncontr')_{[0,k']}
          \models \phi_E$ for all $k'\ge 0$.
        \item Therefore, it has to hold that $\res(s,\Uncontr'')_{[0,k]}
          \models \phi_E$ for all $k\ge 0$.  This in turn implies that
          $\res(s,\Uncontr'')_{[0,k]} \models \phi_C$.  Since
          $\simpleres(s',\Uncontr') = \res(s,\Uncontr'')$, it holds that
          $\simpleres(s',\Uncontr')_{[0,k']} \models \phi_C$ as well, where
          $k'=k$, that is, $\simpleres(s',\Uncontr')_{[0,k']} \models
          \phi_E \to \phi_C$.
      \end{itemize}
    \end{itemize}
  \qed
\end{proof}

Therefore, for checking the realizability of any formula of \DECLARE, one
can check whether the corresponding \LTLf formula in
\cref{lemma:simple:general:real} is realizable with classical algorithms
for \LTLf realizability, like the one proposed in~\cite{DeGiacomoV15}. The
same applies to reactive synthesis as well.

\begin{example}
Considering the two \DECLARE specifications for customer and shop as described in Example~\ref{ex:declare}, one can see that shop can indeed realize a strategy to orchestrate the fragment of the process. Intuitively, the shop would need to wait (i.e., play $skip$) while the
customer opens an order, adds items to it, closes it etc. If the customer pays, the shop needs to wait for the registration of the address unless this was already done. Once this happens, the shop ships. Upon a cancelation request, the shop reacts as follows: if the order has not been shipped yet, it is canceled, otherwise it is partially refunded.
\end{example}

However, since \LTLf realizability is \EXPTIME[2]-complete, such algorithms
work in \emph{doubly exponential time}. This opens the space for the
following question: can we give more efficient algorithms for \DECLARE
realizability?  In the next section, we give a positive answer to this
question, giving an algorithm for \DECLARE realizability that works in
\emph{singly exponential time}.

%% file: sections/example.tex

\newcommand{\respondedexistencepayacc}{\ensuremath{\ltl{F(pay) -> F(acc)}}}

\newcommand{\responsepayget}{\ensuremath{\ltl{G (pay -> XF (get))}}}

\newcommand{\precedencepayget}{pay}

\newcommand{\absencepay}{pay}

\newcommand{\notcoexistencegetcancel}{e}

\newcommand{\nodedist}{6mm}
\newcommand{\taskdist}{9mm}
\newcommand{\autshift}{5mm}

\renewcommand{\taskdist}{2.5cm}
\renewcommand{\arraystretch}{.7}

\definecolor{burntsienna}{rgb}{0.91, 0.45, 0.32}
\definecolor{burntorange}{rgb}{0.8, 0.33, 0.0}
\definecolor{burgundy}{rgb}{0.5, 0.0, 0.13}

\definecolor{darkmagenta}{rgb}{0.55, 0.0, 0.55}
\definecolor{ferngreen}{rgb}{0.31, 0.47, 0.26}

\tikzstyle{custfg} = [
  text=burgundy,
]

\tikzstyle{custbg} = [
  fill=burntsienna!50,
]

\tikzstyle{custline} = [
  draw=burntorange,
]

\tikzstyle{cust} = [
  ultra thick,
  custfg,
  custbg,
  custline
]

\tikzstyle{sellfg} = [
  text=deepblue,
]

\tikzstyle{sellbg} = [
  fill=cyan!50,
]

\tikzstyle{sellline} = [
  draw=deepblue,
]

\tikzstyle{sell} = [
  ultra thick,
  sellfg,
  sellbg,
  sellline
]

\begin{figure}[t]
\centering
\resizebox{.9\textwidth}{!} {
\begin{tikzpicture}[node distance= \taskdist]
  \node[task,cust] (cancel) {
    \begin{tabular}{@{}c@{}}
        \begin{tabular}{@{}c@{}}
        \activity{request}\\
        \activity{cancellation}\\
    \end{tabular}
    \end{tabular}
  };
 
  \node[task,cust, right=3.1cm of cancel] (pay) {
        \activity{pay}
  };
  \node[right=9mm of pay,anchor=east,custfg] (absence2) {\activity{0..1}};
  
  \node[right=16mm of absence2,
          anchor=center,
          draw,
          dotted,
          rounded corners=5pt,
          thick,
          minimum height=6mm]
          (abs) {  \texttt{abscence2}($pay$)  };
  \path[thick,dotted]
    (abs)
    edge
    (absence2);

  \node[right=.45*\taskdist of cancel,
          anchor=center,
          yshift=.43*\taskdist,
          anchor=center,
          draw,
          dotted,
          rounded corners=5pt,
          thick,
          minimum height=6mm]
          (nco) {
      \texttt{neg-succession}($cancel$, $pay$)
  };
  \path[thick,dotted]
    (nco)
    edge
    ($(cancel.east)+(.45*\taskdist,2mm)$);

  \node[task,cust, right=1.88*\taskdist of pay] (register) {
    \begin{tabular}{@{}c@{}}
        \activity{\textbf{register}}\\
        \activity{address}\\
    \end{tabular}
  };

  \node[task,cust, above= of pay,yshift=-6mm] (close) {
          \begin{tabular}{@{}c@{}}
        \activity{\textbf{close}}\\
        \activity{order}\\
    \end{tabular}
  };

     \node[left=.84*\taskdist of close,
          anchor=center,
          yshift=-.3*\taskdist,
          anchor=center,
          draw,
          dotted,
          rounded corners=5pt,
          thick,
          minimum height=6mm]
          (pre1) {
      \texttt{precedence}($close$, $pay$)
  };
    \path[thick,dotted]
    (pre1)
    edge
    ($(pre1.east)+(.34*\taskdist,0)$);

  \node[task,cust, right= .72*\taskdist of close] (add) {
          \begin{tabular}{@{}c@{}}
        \activity{\textbf{add}}\\
        \activity{item}\\
    \end{tabular}
  };

  \node[task,cust, right= .72*\taskdist  of add] (open) {
          \begin{tabular}{@{}c@{}}
        \activity{\textbf{open}}\\
        \activity{order}\\
    \end{tabular}
  };

   \node[below=.55*\taskdist of open,
          anchor=center,
          xshift=-1.15*\taskdist,
          anchor=center,
          draw,
          dotted,
          rounded corners=5pt,
          thick,
          minimum height=6mm]
          (resp1) {
      \texttt{resp-existence}($open$, $regAdd$)     
  };
  \path[thick,dotted]
    (resp1)
    edge
    ($(resp1.east)+(.5,0)$);

  \draw[negationsuccession,cust] (pay) -- (cancel);
  \draw[precedence,cust] (pay) -- (close);
  \draw[precedence,cust,transform canvas={yshift=2.5mm}] (close) -- (add);
  \draw[negationsuccession,cust,transform canvas={yshift=-2.5mm}] ( add) -- ( close);
  \draw[precedence,cust] (add) -- (open);
  \draw[precedence,cust] (pay) -- (close);
   \draw[respondedexistence,cust] (open) -- (register);


\node[task,sell, below= .7*\taskdist  of cancel,xshift=-1.8cm] (refund) {
          \begin{tabular}{@{}c@{}}
        \activity{\textbf{partial}}\\
        \activity{refund}\\
    \end{tabular}
  };  

\node[task,sell, below= .7*\taskdist  of cancel,xshift=1.8cm] (cancelo) {
          \begin{tabular}{@{}c@{}}
        \activity{\textbf{cancel}}\\
        \activity{order}\\
    \end{tabular}
  };  
  
  \draw[notcoexistence,sell] (refund) -- (cancelo);

  \node[left=.05*\taskdist of cancelo,
          anchor=center,
          yshift=-.45*\taskdist,
          anchor=center,
          draw,
          dotted,
          rounded corners=5pt,
          thick,
          minimum height=6mm]
          (nco1) {
      \texttt{not-coexistence}($cancel$, $refund$)
  };
    \path[thick,dotted,transform canvas={shift={(-1,0)}}]
    (nco1)
    edge
    ($(nco1.north)+(0,.6)$);

  \node[task,sell, below= .7*\taskdist  of pay] (ship) {
          \begin{tabular}{@{}c@{}}
        \activity{ship}\\
        \activity{order}\\
    \end{tabular}
  };  
  
  \draw[negationsuccession,sell] (cancelo) -- (ship);

  
   \node[task,sell, right= 2*\taskdist  of ship] (skip) { \activity{skip} };

  \draw[precedence,sellline] (refund.north east) to [bend left= 23] (ship.north west);

  \draw[respondedexistence,darkmagenta] (cancel) -- ($(cancel.south)-(0,.6)$) -- (-1.8,-1.8);
  \draw[respondedexistence,darkmagenta] (cancel) -- ($(cancel.south)-(0,.6)$) -- (1.8,-1.8);
  \draw[constraint,-triangle 60,darkmagenta] (1.8,-1.78) -- (cancelo.north);
  \draw[constraint,-triangle 60,darkmagenta] (-1.8,-1.78) -- (refund.north);
  \draw[succession, darkmagenta] (ship) -- (pay);
  \draw[precedence, darkmagenta] (ship) to [out=0,in=-90] (register);
  \draw[negationsuccession, ferngreen] (register) to [out=240,in=20] (ship);

    \node[right=.6*\taskdist of pay,
          anchor=center,
          yshift=-.46*\taskdist,
          anchor=center,
          draw,
          dotted,
          rounded corners=5pt,
          thick,
          minimum height=6mm]
          (suc) {
      \texttt{succession}($pay$, $ship$)
  };
    \path[thick,dotted]
    (suc)
    edge
    ($(suc.west)-(.4,0)$);

 \node[right=.6*\taskdist of cancel,
          anchor=center,
          yshift=-.4*\taskdist,
          anchor=center,
          draw,
          dotted,
          rounded corners=5pt,
          thick,
          minimum height=6mm]
          (resp) {
           \begin{tabular}{@{}r@{}}
       \texttt{response}($reqCancel$, \\
        $refund\lor cancel$)\\
    \end{tabular}
     
  };
    \path[thick,dotted]
    (resp)
    edge
    ($(resp.west)-(.93,0)$);

\end{tikzpicture}
}
\caption{A graphical representation of an order handling process in
\DECLARE. Actions are depicted with rectangles and the arrows denote \DECLARE constraints, some of which come with additional (descriptive) annotations. 
The orange color highlights the customer-related actions and constraints, blue does the same for the seller. 
In magenta and green we put the constraints needed to link behaviours of the players. 
\label{fig:declare}}
\end{figure}
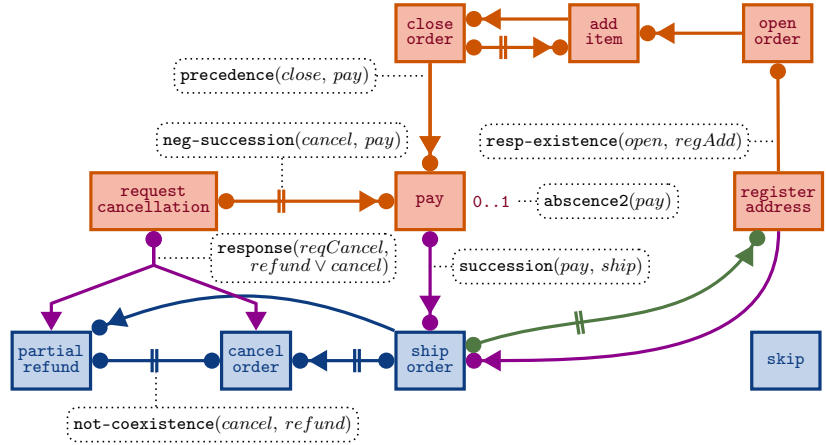

%% file: sections/4.explicit.tex

\section{Efficient Reactive Synthesis for \DECLARE}
\label{sec:explicit}

In this section, we give an algorithm for the realizability problem of
\DECLARE that works in singly exponential time.
It is based on two steps:
\begin{enumerate*}[label=(\roman*)]
  \item a transformation of any \DECLARE model into an equivalent
    \emph{pure past} formula;
  \item the construction of an equivalent singly-exponential \DFA (recall
    \cref{prop:ltlfp:dfa}), that can be used as an arena for a reachability
    game.
\end{enumerate*}


\subsection{Pastification for \DECLARE}

The first step of our algorithm is called \emph{pastification}, and
consists of transforming each \DECLARE model into an equivalent one that
uses only past temporal operators.
In this section, we give a pastification algorithm for any \DECLARE pattern
and any \DECLARE model. We begin with the definition of
\emph{pastification}.

Pastification~\cite{maler2007synthesizing,cimatti2021extended} is
a technique to transform temporal formulas into equivalent pure past ones (\ie, those with past operators only).
The main advantage of pure past formulas is
that, since ``past already happened'', there is no need for nondeterminism,
and so such formulas can be transformed into deterministic automata
with only a singly exponential blowup (recall \cref{prop:ltlfp:dfa}), thus
more efficiently with respect to general \LTL.
In this section, we exploit this fundamental feature to obtain \emph{singly
exponential-size} \DFAs starting from any \DECLARE model.
We define the pastification for \DECLARE models as follows.
\begin{definition}[Pastification for \DECLARE models]
\label{def:declare:pastification}
  Let $\phi$ be a \DECLARE model over the alphabet $\Sigma$.
  A \emph{pastification of $\phi$}, denoted as $\pastify(\phi)$, is
  a formula of \LTLfFP over $\Sigma$ such that, for any finite trace
  $\sigma\in(2^\Sigma)^+$, it holds that:
  \begin{align}
    \sigma,0 \models\phi \ \Iff \ \sigma,|\sigma|-1 \models \pastify(\phi)
  \end{align}
\end{definition}

From now until the end of this section, without loss of generality we
assume general finite traces semantics.

\paragraph{Pastification of \DECLARE patterns.}
In the third column of \cref{tab:constraints}, we list, for each \DECLARE
pattern $\phi$, the corresponding pastification $\pastify(\phi)$.  The
following theorem establishes that, for any \DECLARE pattern $\phi$,
$\pastify(\phi)$ is a pastification of $\phi$ of size linear in $|\phi|$.

\begin{restatable}{theorem}{thpastcorrectness}
\label{th:past:correctness}
  For each pattern in \cref{tab:constraints}, let $\phi$ be the \LTLf
  formula in the second column and let $\psi$ be the formula in the third
  column. It holds that $\psi$ is a pastification of $\phi$ and $|\psi| \in
  \mathcal{O}(|\phi|)$.
\end{restatable}
\begin{proof}
  For any pattern in \cref{tab:constraints} where $\phi$ and $\psi$ are the
  formulas in the second and third column, respectively, we prove that
  \begin{enumerate*}[label=(\roman*)]
    \item $\sigma \models \phi$ iff $\sigma,|\sigma|-1 \models \psi$, for
      any $\sigma \in (2^\Sigma)^*$, and
    \item $|\psi| \in \mathcal{O}(|\phi|)$.
  \end{enumerate*}
  We prove the first point. For each \DECLARE pattern $\phi$, we show that
  $\psi$ is a pastification of $\phi$.

  \paragraph{The \texttt{existence}($p$) pattern.}
  For any $\sigma\in(2^\Sigma)^*$, it holds that:
  \begin{align}
    &\sigma,0 \models \ltl{F(p)} \\
    \Iff \ 
    &\exists 0\le i < |\sigma| \suchdot (p \in \sigma_i) \\
    \Iff \ 
    &\sigma,|\sigma|-1 \models \ltl{O(p)}
  \end{align}

  \paragraph{The \texttt{absence2}($p$) pattern.}
  For any $\sigma\in(2^\Sigma)^*$, it holds that:
  \begin{align}
    &\sigma,0 \models \ltl{! F(p & X F(p))} \\
    \Iff \ 
    &\lnot \exists 0\le i < |\sigma| \suchdot (p \in \sigma_i \land
     \exists i< j < |\sigma| \suchdot p \in \sigma_j)\\
    \Iff \ 
    &\lnot \exists 0\le i < |\sigma| \suchdot (p \in \sigma_i \land
     \exists 0< j < i \suchdot p \in \sigma_j)\\
    \Iff \ 
    &\forall 0\le i < |\sigma| \suchdot (p \in \sigma_i \to
     \forall 0< j < i \suchdot p \not\in \sigma_j)\\
    \Iff \ 
    &\sigma,|\sigma|-1 \models \ltl{H(p \to Z H (!p))}
  \end{align}

  \paragraph{The \texttt{choice}($p,q$) pattern.}
  For any $\sigma\in(2^\Sigma)^*$, it holds that:
  \begin{align}
    &\sigma,0 \models \ltl{F(p) | F(q)} \\
    \Iff \ 
    &\exists 0\le i < |\sigma| \suchdot (p\in\sigma_i) \lor
     \exists 0\le i < |\sigma| \suchdot (q\in\sigma_i)\\
    \Iff \ 
    &\exists 0\le i < |\sigma| \suchdot (p\in\sigma_i \lor q\in\sigma_i)\\
    \Iff \ 
    &\sigma,|\sigma|-1 \models \ltl{O(p | q)}
  \end{align}

  \paragraph{The \texttt{exc-choice}($p,q$) pattern.}
  For any $\sigma\in(2^\Sigma)^*$, it holds that:
  \begin{align}
    &\sigma,0 \models \ltl{(F(p) | F(q)) & !(F(p) & F(q))} \\
    \Iff \ 
    &(\exists 0\le i < |\sigma| \suchdot (p \in \sigma_i) \lor
     \exists 0\le i < |\sigma| \suchdot (q \in \sigma_i)) \land \\
     &\lnot (\exists 0 \le i < |\sigma| \suchdot (p \in \sigma_i) \land
            \exists 0 \le i < |\sigma| \suchdot (q \in \sigma_i)) \\
    \Iff \ 
    &(\exists 0\le i < |\sigma| \suchdot (p \in \sigma_i \lor q \in \sigma_i)) \land \\
     &\forall 0 \le i < |\sigma| \suchdot (p \not\in \sigma_i) \lor
     \forall 0 \le i < |\sigma| \suchdot (q \not\in \sigma_i) \\
    \Iff \ 
    &\sigma,|\sigma|-1 \models \ltl{O(p | q) & (H(!p) | H(!q))}
  \end{align}

  \paragraph{The \texttt{resp-existence}($p$) pattern.}
  For any $\sigma\in(2^\Sigma)^*$, it holds that:
  \begin{align}
    &\sigma,0 \models \ltl{F(p) -> F(q)} \\
    \Iff \ 
    &\exists 0\le i < |\sigma| \suchdot (p \in \sigma_i) \to
     \exists 0\le i < |\sigma| \suchdot (q \in \sigma_i)\\
    \Iff \ 
    &\forall 0\le i < |\sigma| \suchdot (p \not\in \sigma_i) \lor
     \exists 0\le i < |\sigma| \suchdot (q \in \sigma_i)\\
    \Iff \ 
    &\sigma,|\sigma|-1 \models \ltl{H(!p) | O(q)}
  \end{align}

  \paragraph{The \texttt{coexistence}($p,q$) pattern.}
  For any $\sigma\in(2^\Sigma)^*$, it holds that:
  \begin{align}
    &\sigma,0 \models \ltl{F(p) \iff F(q)} \\
    \Iff \ 
    &\sigma,0 \models \ltl{(F(p) \to F(q)) & (F(q) \to F(p))} \\
    \Iff \ 
    &\sigma,0 \models \mathtt{resp{\text -}existence}(p,q) \land \mathtt{resp{\text -}existence}(q,p) \\
    \Iff \ 
    &\sigma,|\sigma|-1 \models \ltl{(H(!p) | O(q)) & (H(!q) | O(p))}
  \end{align}

  \paragraph{The \texttt{response}($p,q$) pattern.}
  For any $\sigma\in(2^\Sigma)^*$, it holds that:
  \begin{align}
    &\sigma,0 \models \ltl{G(p -> F(q))} \\
    \Iff \ 
    &\forall 0\le i<|\sigma| \suchdot (p\in\sigma_i \to \exists i\le
     j<|\sigma| \suchdot q\in\sigma_j) \\
    \Iff \ 
    &\lnot\exists 0\le i<|\sigma| \suchdot (p\in\sigma_i \land
     q\not\in\sigma_i \land \forall i< j < |\sigma| \suchdot (q
     \not\in\sigma_j)) \\
    \Iff \ 
    &\sigma,|\sigma|-1 \models \ltl{!((!q) S (p & !q))} \\
    \Iff \ 
    &\sigma,|\sigma|-1 \models \ltl{q T (!p | q)}
  \end{align}

  \paragraph{The \texttt{precedence}($p,q$) pattern.}
  For any $\sigma\in(2^\Sigma)^*$, it holds that:
  \begin{align}
    &\sigma,0 \models \ltl{(! q) W (p)} \\
    \Iff \ 
    &\exists 0\le i<|\sigma| \suchdot (p \in \sigma_i \land \forall 0 \le
     j< i \suchdot (q \not\in \sigma_j)) \lor \\
     &\forall 0 \le i<|\sigma| \suchdot (q \not\in \sigma_i) \\
    \Iff \ 
    &\lnot\exists 0\le i<|\sigma| \suchdot (q \in \sigma_i \land \forall 0 \le
     j< i \suchdot (p \not\in \sigma_j)) \\
    \Iff \ 
    &\forall 0\le i < |\sigma| \suchdot (q\in\sigma_i \to \exists 0\le
     j\le i \suchdot (p\in\sigma_j)) \\
    \Iff \ 
    &\sigma,|\sigma|-1 \models \ltl{H(q -> O(p))}
  \end{align}

  \paragraph{The \texttt{succession}($p,q$) pattern.}
  For any $\sigma\in(2^\Sigma)^*$, it holds that:
  \begin{align}
    &\sigma,0 \models \ltl{G(p -> F(q)) & (! q) W (p)} \\
    \Iff \ 
    &\sigma,0 \models \mathtt{response}(p,q) \land \mathtt{precedence}(p,q) \\
    \Iff \ 
    &\sigma,|\sigma|-1 \models \ltl{p T (!p | q) & H(q -> O(p))}
  \end{align}

  \paragraph{The \texttt{alt-response}($p,q$) pattern.}
  For any $\sigma\in(2^\Sigma)^*$, it holds that:
  \begin{align}
    &\sigma,0 \models \ltl{G(p -> X((! p) U q))} \\
    \Iff \ 
    &\forall 0\le i<|\sigma| \suchdot (p\in\sigma_i \to \\
     &\exists
     i<j<|\sigma| \suchdot (q\in\sigma_j \land \forall i<k<j \suchdot
     (p\not\in\sigma_k))) \\
    \Iff \ 
    &\lnot\left(
       \begin{aligned}
         &\exists 0\le i<|\sigma|\suchdot (p\in\sigma_i \land \forall
         i<j<|\sigma| \suchdot (p\not\in\sigma_j \land q\not\in\sigma_j))
           \lor \\
           &\left(\begin{aligned}
             &\exists 0\le i<|\sigma| \suchdot (i>0 \land q\in\sigma_i \land \exists
             0\le j < i \suchdot ( p\in\sigma_j \land q\not\in\sigma_j \land \\
             &\exists 0\le h<j\suchdot (p\in\sigma_h \land \forall h<l<j \suchdot
             q\not\in\sigma_l) \land \forall j<k<i \suchdot (q\not\in\sigma_k)))
           \end{aligned}
           \right)
       \end{aligned}
    \right) \\
    \Iff \ 
     &\sigma,0\models\lnot \left(
       \begin{aligned}
         &\ltl{(!q & !p) S (p)} \ \lor \\
         &\ltl{O(q \land Y((!q)S(p & !q & Y((!q)S(p)))))}
       \end{aligned}
     \right) \\
    \Iff \ 
    &\sigma,|\sigma|-1 \models \ltl{(p | q) T (!p) & H(q -> wY(q T ((p & !q) -> Z(q T !p))))}
  \end{align}

  \paragraph{The \texttt{alt-precedence}($p,q$) pattern.}
  For any $\sigma\in(2^\Sigma)^*$, it holds that:
  \begin{align}
    &\sigma,0 \models \ltl{((! q) W p) & G(q -> wX((! q) W p))} \\
    \Iff \ 
    &\left(
     \begin{aligned}
       &\exists 0\le i<|\sigma| \suchdot (p\in\sigma_i \land \forall 0\le
        j < i \suchdot (q\not\in\sigma_j)) \lor \forall 0\le i<|\sigma|
        \suchdot (q\not\in\sigma_i)) 
        \land \\
       &\forall 0\le i<|\sigma| \suchdot (q\in\sigma_i \to \left( 
        \begin{aligned}
          &\exists i < j < |\sigma| \suchdot (p \in \sigma_j \land \forall
          i<k<j \suchdot (q\not\in\sigma_k)) \lor \\
          &\forall i<j<|\sigma|
          \suchdot (q\not\in\sigma_j)
        \end{aligned}
        \right))
     \end{aligned}
     \right)
    \\
    \Iff \ 
    &\lnot\left(
     \begin{aligned}
       &\exists 0\le i<|\sigma| \suchdot (q\in\sigma_i \land \forall 0\le
      j \le i \suchdot (p\not\in\sigma_j))
      \lor \\
       &\exists 0\le i<|\sigma| \suchdot \left(
        \begin{aligned}
          &i>0 \land q\in\sigma_i \land p\not\in\sigma_i \land \\
          &\exists 0\le j < i \suchdot ( q\in\sigma_j \land \\
          &\exists 0\le h<j \suchdot (p\in\sigma_j \land \forall h<l\le
          j \suchdot (p\not\in\sigma_l)) \land \\
          &\forall j<k<i \suchdot (p\not\in\sigma_k))
        \end{aligned}
       \right)
     \end{aligned}
    \right) \\
    \Iff \ 
    &\sigma,|\sigma|-1 \models \lnot(
      \ltl{O(q & H(!p))}
      \lor
      \ltl{O((q & !p) & Y((!p)S(q & (!p)S(p))))}
    ) \\
    \Iff \ 
    &\sigma,|\sigma|-1 \models \ltl{H(q -> O(p)) & H((q & !p) -> wY (p T (q -> (p T (! p)))))}
  \end{align}

  \paragraph{The \texttt{alt-succession}($p,q$) pattern.}
  For any $\sigma\in(2^\Sigma)^*$, it holds that:
  \begin{align}
    &\sigma,0 \models \mathtt{alt{\text -}response}(p,q) \land \mathtt{alt{\text -}precedence}(p,q) \\
    \Iff \ 
    &\sigma,|\sigma|-1 \models \pastify(\mathtt{alt{\text -}response}(p,q))
    \land \pastify(\mathtt{alt{\text -}precedence}(p,q))
  \end{align}

  \paragraph{The \texttt{chain-response}($p,q$) pattern.}
  For any $\sigma\in(2^\Sigma)^*$, it holds that:
  \begin{align}
    &\sigma,0 \models \ltl{G(p -> X(q))} \\
    \Iff \ 
    &\forall 0\le i<|\sigma| \suchdot (p\in\sigma_i \to ( i<|\sigma|-1
    \land q\in\sigma_{i+1}) \\
    \Iff \ 
    &\forall 0\le i<|\sigma| \suchdot ((i>0 \land p\in\sigma_{i-1}) \to
    q\in\sigma_{i}) \land p\not\in\sigma_{|\sigma|-1} \\
    \Iff \ 
    &\sigma,|\sigma|-1 \models \ltl{! p & H(Y(p) -> q)}
  \end{align}

  \paragraph{The \texttt{chain-precedence}($p,q$) pattern.}
  For any $\sigma\in(2^\Sigma)^*$, it holds that:
  \begin{align}
    &\sigma,0 \models \ltl{G(X(q) -> p)} \\
    \Iff \ 
    &\forall 0\le i<|\sigma| \suchdot ((i<|\sigma|-1 \land
     q\in\sigma_{i+1}) \to p\in\sigma_{i}) \\
    \Iff \ 
    &\forall 0\le i<|\sigma| \suchdot (q\in\sigma_i \to (i=0 \lor
     p\in\sigma_{i-1})) \\
    \Iff \ 
    &\sigma,|\sigma|-1 \models \ltl{H(q -> wY p)}
  \end{align}

  \paragraph{The \texttt{chain-succession}($p,q$) pattern.}
  For any $\sigma\in(2^\Sigma)^*$, it holds that:
  \begin{align}
    &\sigma,0 \models \ltl{G(p \iff X(q))} \\
    \Iff \ 
    &\sigma,0 \models \mathtt{chain{\text -}response}(p,q) \land
    \mathtt{chain{\text -}precedence}(p,q) \\
    \Iff \ 
    &\sigma,|\sigma|-1 \models \pastify(\mathtt{chain{\text
    -}response}(p,q)) \land \pastify(\mathtt{chain{\text
    -}precedence}(p,q))
  \end{align}

  \paragraph{The \texttt{not-coexistence}($p,q$) pattern.}
  For any $\sigma\in(2^\Sigma)^*$, it holds that:
  \begin{align}
    &\sigma,0 \models \ltl{!(F(p) & F(q))} \\
    \Iff \ 
    &\lnot(\exists 0\le i<|\sigma| \suchdot(p\in\sigma_i) \land \exists
    0\le i<|\sigma| \suchdot (q\in\sigma_i)) \\
    \Iff \ 
    &\forall 0\le i<|\sigma| \suchdot(p\not\in\sigma_i) \lor \forall 0\le
    i<|\sigma| \suchdot (q\not\in\sigma_i) \\
    \Iff \ 
    &\sigma,|\sigma|-1 \models \ltl{H(!p) | H(!q)}
  \end{align}

  \paragraph{The \texttt{neg-succession}($p,q$) pattern.}
  For any $\sigma\in(2^\Sigma)^*$, it holds that:
  \begin{align}
    &\sigma,0 \models \ltl{G(p -> ! F(q))} \\
    \Iff \ 
    &\forall 0\le i<|\sigma| \suchdot (p\in\sigma_i \to \lnot \exists i\le
    j < |\sigma| \suchdot (q\in\sigma_j)) \\
    \Iff \ 
    &\left(
    \begin{aligned}
      &\forall 0\le i<|\sigma| \suchdot (p\not\in\sigma_i)
      \lor \\
      &\exists 0\le i<|\sigma| \suchdot (p\in\sigma_i \land
      q\not\in\sigma_i \land \forall 0\le j<i \suchdot (p\not\in\sigma_j)
      \land \\ 
      &\forall i<j<|\sigma| \suchdot (q\not\in\sigma_j))
    \end{aligned}
    \right)\\
    \Iff \ 
    &\sigma,|\sigma|-1 \models \ltl{H(!p) | (!q)S(p & !q & wY H(!p))}
  \end{align}

  \paragraph{The \texttt{neg-chain-succession}($p,q$) pattern.}
  For any $\sigma\in(2^\Sigma)^*$, it holds that:
  \begin{align}
    &\sigma,0 \models \ltl{G(p -> wX(! q)) \land G(q -> wX(! p))} \\
    \Iff \ 
    &\forall 0\le i<|\sigma| \suchdot (p\in\sigma_i \to (i=|\sigma|-1 \lor
    q\not\in\sigma_{i+1})) \land \\
    &\forall 0\le i<|\sigma| \suchdot
    (q\in\sigma_i \to (i=|\sigma|-1 \lor p\not\in\sigma_{i+1})) \\
    \Iff \ 
    &\forall 0\le i<|\sigma| \suchdot ((i>0 \land p\in\sigma_{i-1}) \to
    q\not\in\sigma_{i})) \land \\
    &\forall 0\le i<|\sigma| \suchdot
    ((i>0 \land q\in\sigma_{i-1} \to p\not\in\sigma_{i}) \\
    \Iff \ 
    &\sigma,|\sigma|-1 \models \ltl{H(Y(p) -> ! q) \land H(Y(q) -> ! p)}
  \end{align}

  Finally, it is straightforward to see that, for each $\phi$ and $\psi$,
  the size of $\psi$ is linear in the size of $\phi$.
  \qed
\end{proof}

It is worth pointing out the following property of
\cref{def:declare:pastification}: since the formula $\ltl{wX(\false)}$ is
true iff it is interpreted at the last time point of a (finite) trace,
\cref{def:declare:pastification} amounts to require that the formula $\phi
\iff \ltl{F(wX(\false) & \pastify(\phi))}$ is \emph{valid}.  As a matter of
fact, we checked the correctness of the pastification of all \DECLARE
patterns by checking the validity\footnotemark of formulas of the previous
type with the \BLACK tool~\cite{GeattiGMV21,GeattiGM21}.
\footnotetext{
  Since \BLACK is a tool for satisfiability checking of \LTL and \LTLf (and
  related extensions), we reduced the validity checking of a formula $\psi$
  to the satisfiability checking of the formula $\lnot\phi$ ($\psi$ is
  valid iff $\lnot\phi$ is unsatisfiable).
}

We can generalize \Cref{th:past:correctness} (that pastifies each pattern
of \DECLARE) to any \DECLARE model, \ie to any conjunction of patterns. 

\begin{restatable}{corollary}{corpastcorrectness}
\label{cor:past:correctness}
  For any \DECLARE model $\phi$, there exists a formula $\psi$ of \LTLfFP
  such that $\psi$ is a pastification of $\phi$ and
  $|\psi|\in\mathcal{O}(|\phi|)$.
\end{restatable}
\begin{proof}
  By definition, $\phi$ is a conjunction $\phi_1 \land \dots \land \phi_n$
  of \DECLARE patterns (for some $n\in\N$). By \cref{th:past:correctness},
  there exist $\psi_1,\dots,\psi_n$ such that $\psi_i \in \LTLfFP$,
  $\psi_i$ is a pastification of $\phi_i$, and
  $|\psi_i|\in\mathcal{O}(|\phi_i|)$, for each $1\le i\le n$. Let $\psi
  \coloneqq \psi_1 \land \dots \land \psi_n$.  Clearly, it holds that
  $|\psi| \in \mathcal{O}(n)$. Moreover, for any (finite) state sequence
  $\sigma$, we have that $\sigma,0 \models \phi$ iff $\sigma,0 \models
  \phi_i$ (for each $1\le i\le n$) iff $\sigma,|\sigma|-1 \models \psi_i$
  (for each $1\le i\le n$) iff $\sigma,|\sigma|-1 \models \psi$. It follows
  that $\psi$ is a pastification of $\phi$.
  \qed
\end{proof}

One can wonder whether \cref{th:past:correctness,cor:past:correctness} can
be generalized to any formula of full \LTLf. We show that this is
impossible, \ie a polynomial-size pastification for full \LTLf cannot
exists.  Suppose by contradiction that this holds. Then, for any formula
$\phi$ of \LTLf we apply such pastification algorithm, obtaining a formula
$\psi$ of \LTLfFP that is a pastification of $\phi$ and is of size
polynomial in $\phi$. By the property of being
a pastification (see \cref{def:declare:pastification}), $\phi$ is
realizable iff $\psi$ is realizable. Since $\psi$ is a formula of \LTLfFP
and realizability of \LTLfFP is a \EXPTIME-complete problem~\cite{safetycomplexity},
it follows that also the realizability problem of \LTLf is
\EXPTIME-complete, which is a contradiction, since it is
\EXPTIME[2]-complete~\cite{DeGiacomoV15}.
\Cref{th:past:correctness} shows that \DECLARE is a fragment of \LTLf that
admits a polynomial-space pastification.

\paragraph{Pastification of auxiliary formulas.}
In order to reduce \DECLARE realizability to \LTLf realizability according
to \cref{lemma:simple:general:real}, in the following lemma we give
a (polynomial-size) pastification of the formulas
$\mathtt{simple}_{\mathtt{Env}}(\Uset)$ and
$\mathtt{simple}_{\mathtt{Con}}(\Cset)$.
\begin{restatable}{lemma}{propsimplepast}
\label{prop:simple:past}
  There exists two linear-size pastifications of
  $\mathtt{simple}_{\mathtt{Env}}(\Uset)$ and \\
  $\mathtt{simple}_{\mathtt{Con}}(\Cset)$.
\end{restatable}
\begin{proof}
  We prove that the formula:
  \begin{align}
    \ltl{O}(\ltl{Z} \bot \land \bigvee_{u\in\Uset} u) \land \ltl{H}\left(
    \begin{aligned}
      &(\bigvee_{u\in\Uset}u \to \bigwedge_{u\not=u'\in\Uset}\lnot (u \land
      u')) \ \land \\
      &(\ltl{Y}(\bigvee_{u\in\Uset}u) \to \bigwedge_{u\in\Uset}\lnot u)
      \land (\ltl{Y Y}(\bigvee_{u\in\Uset} u) \to \bigvee_{u\in\Uset} u)
    \end{aligned}
    \right)
  \end{align}
  and the formula
  \begin{align}
    \ltl{O}(\ltl{Z} \bot \land \bigwedge_{c\in\Cset}\lnot c)\land\ltl{H}\left(
    \begin{aligned}
      &(\ltl{Y}(\bigwedge_{c\in\Cset} \lnot c) \to (\bigvee_{c\in\Cset} c \land
      \bigwedge_{c\not=c'\in\Cset} \lnot (c \land c'))) \land \\
      &(\ltl{Y Y}(\bigwedge_{c\in\Cset}\lnot c)) \to (\bigwedge_{c\in\Cset}
      \lnot c)
    \end{aligned}
    \right)
  \end{align}
  are linear-size pastifications of $\mathtt{simple}_{\mathtt{Env}}(\Uset)$
  and $\mathtt{simple}_{\mathtt{Con}}(\Cset)$ , respectively.

  We start proving the result for the first formula.  Suppose that, for any
  trace $\sigma\in(2^\Sigma)^+$, it holds that:
  \begin{align}
    &\sigma,0 \models
      \bigvee_{u\in\Uset}u \land \ltl{G}(\bigvee_{u\in\Uset}u \to
      (\bigwedge_{u\not=u'\in\Uset} \lnot(u \land u') \land
      \ltl{wX}(\bigwedge_{u\in\Uset} \lnot u \land \ltl{wX}
      \bigvee_{u\in\Uset} u)))
  \end{align}
  By definition of the temporal operators, this amounts to:
  \begin{align}
    &\bigvee_{u\in\Uset}u\in\sigma_0 \land \forall 0 \le j<|\sigma|
    \suchdot \left(
    \begin{aligned}
      &((\bigvee_{u\in\Uset}u\in\sigma_j) \to
      (\bigwedge_{u\not=u'\in\Uset} \lnot(u\in\sigma_j \land u'\in\sigma_j)
      \land \\
      &(j=|\sigma|-1 \lor ((\bigwedge_{u\in\Uset} \lnot u\in\sigma_{j+1})
      \land \\
      &(j=|\sigma|-2 \lor (\bigvee_{u\in\Uset} u\in\sigma_{j+2}))))))
    \end{aligned}
    \right)
  \end{align}
  This is equivalent to:
  \begin{align}
    &\sigma,|\sigma|-1 \models 
      \exists 0 \le i < |\sigma| \suchdot (i=0 \land \bigvee_{u\in\Uset}
      u\in\sigma_i) \land \forall 0\le i < |\sigma| \suchdot \\
      &\left(
      \begin{aligned}
        &(\bigvee_{u\in\Uset}u\in\sigma_i \to
        \bigwedge_{u\not=u'\in\Uset}\lnot (u\in\sigma_i \land
        u'\in\sigma_i)) \ \land \\
        &((i>0 \land \bigvee_{u\in\Uset}u\in\sigma_i) \to
        \bigwedge_{u\in\Uset}u\not\in\sigma_i)
        \land ((i>1 \land \bigvee_{u\in\Uset} u\in\sigma_i) \to
        \bigvee_{u\in\Uset} u\in\sigma_i)
      \end{aligned}
      \right)
  \end{align}
  That is:
  \begin{align}
    &\sigma,|\sigma|-1 \models 
      \ltl{O}(\ltl{Z} \bot \land \bigvee_{u\in\Uset} u) \land \ltl{H}\left(
      \begin{aligned}
        &(\bigvee_{u\in\Uset}u \to \bigwedge_{u\not=u'\in\Uset}\lnot (u \land
        u')) \ \land \\
        &(\ltl{Y}(\bigvee_{u\in\Uset}u) \to \bigwedge_{u\in\Uset}\lnot u)
        \land (\ltl{Y Y}(\bigvee_{u\in\Uset} u) \to \bigvee_{u\in\Uset} u)
      \end{aligned}
      \right)
  \end{align}

  We now prove the result for the second formula.
  Suppose that, for any trace $\sigma\in(2^\Sigma)^+$, it holds that:
  \begin{align}
    &\sigma,0\models
    \bigwedge_{c\in\Cset}\lnot c \land \ltl{G}(\bigwedge_{c\in\Cset}\lnot
    c \to \ltl{wX}(\bigvee_{c\in\Cset} c \land \bigwedge_{c\not=c'\in\Cset}
    \lnot (c \land c') \land \ltl{wX} \bigwedge_{c\in\Cset} \lnot c)))
  \end{align}
  By the definition of temporal operators, this amounts to:
  \begin{align}
    &\exists 0\le i<|\sigma|\suchdot(\bigwedge_{c\in\Cset}c\not\in\sigma_i)
    \land \forall 0\le i<|\sigma| \suchdot \\
    &\left(
    \begin{aligned}
      &\bigwedge_{c\in\Cset} c\not\in\sigma_i \to (i<|\sigma|-1 \land
      \bigvee_{c\in\Cset} c\in\sigma_{i+1} \land \\
      &\bigwedge_{c\not=c'\in\Cset} \lnot (c\in\sigma_{i+1} \land
      c'\in\sigma_{i+1}) \land (i<|\sigma|-2 \land \bigwedge_{c\in\Cset}
      c\not\in\sigma_{i-2}))
    \end{aligned}
    \right)
  \end{align}
  This is equivalent to:
  \begin{align}
    &\sigma,|\sigma|-1\models \exists 0\le i<|\sigma| \suchdot (i=0 \land
    \bigwedge_{c\in\Cset}c\not\in\sigma_i) \land \forall 0\le i<|\sigma|
    \suchdot \\
    &\left(
    \begin{aligned}
      &((i>0 \bigwedge_{c\in\Cset} c\not\in\sigma_{i-1}) \to
      (\bigvee_{c\in\Cset} c\in\sigma_{i-1} \land
      \bigwedge_{c\not=c'\in\Cset} \lnot (c\in\sigma_{i-1} \land
      c'\in\sigma_{i-1}))) \land \\
      &((i>1 \land \bigwedge_{c\in\Cset}c\not\in\sigma_{i-2})) \to
      (\bigwedge_{c\in\Cset} c\not\in\sigma_{i-2})
    \end{aligned}
    \right)
  \end{align}
  which in turn is equivalent to:
  \begin{align}
    &\sigma,|\sigma|-1\models
    \ltl{O}(\ltl{Z} \bot \land \bigwedge_{c\in\Cset}\lnot c)\land\ltl{H}\left(
    \begin{aligned}
      &(\ltl{Y}(\bigwedge_{c\in\Cset} \lnot c) \to (\bigvee_{c\in\Cset} c \land
      \bigwedge_{c\not=c'\in\Cset} \lnot (c \land c'))) \land \\
      &(\ltl{Y Y}(\bigwedge_{c\in\Cset}\lnot c)) \to (\bigwedge_{c\in\Cset}
      \lnot c)
    \end{aligned}
    \right)
  \end{align}
  Finally, it is straightforward to see that both pastifications are of
  size linear in the size of the original formulas.
  \qed
\end{proof}

\subsection{A singly exponential time algorithm for \DECLARE realizability}
\label{sub:efficient}

The linear-size pastification of \DECLARE patterns and of the
$\mathtt{simple}_{\mathtt{Env}}(\Uset)$ and
$\mathtt{simple}_{\mathtt{Con}}(\Cset)$ formulas allow for the
following singly exponential time algorithm for solving \DECLARE
realizability.

Let $\phi_E$ and $\phi_C$ be two \DECLARE models for the assumptions on the
environment and the guarantees of the controller, respectively.
By \cref{lemma:simple:general:real}, $(\phi_E,\phi_C)$ is realizable iff
the \LTLf formula $\Gamma \coloneqq \mathtt{simple}_{\mathtt{Con}}(\Cset)
\land ((\mathtt{simple}_{\mathtt{Env}}(\Uset) \land \phi_E) \to \phi_C)$ is
realizable.
Now, let $\psi_E$, $\psi_C$, $\psi_{\mathtt{SimpleEnv}}$ and
$\psi_{\mathtt{SimpleCon}}$ be linear-size pastifications of $\phi_E$,
$\phi_C$, $\mathtt{simple}_{\mathtt{Env}}(\Uset)$ and
$\mathtt{simple}_{\mathtt{Con}}(\Cset)$, respectively (by
\cref{cor:past:correctness,prop:simple:past}, they are guaranteed to
exist). Let $\Gamma_P$ be the \emph{pure past} formula
$\psi_{\mathtt{SimpleCon}} \land ((\psi_{\mathtt{SimpleEnv}} \land \psi_E)
\to \psi_C)$.
By definition of pastification, the language of $\Gamma$ is equivalent to
the language of $\Gamma_P$\footnotemark.
\footnotetext{%
 Being a pure past formula, the language of $\Gamma_P$ is
  defined as the set of finite traces that are models of $\Gamma_P$ when it
  is interpreted at the last time point of the trace.
}
By \cref{prop:ltlfp:dfa}, we can effectively build a \DFA $\autom$ such
that $\lang(\autom) = \lang(\Gamma_P)$ and $|\autom| \in
2^{\mathcal{O}(|\Gamma_P|)}$, \ie the size of $\autom$ is at most
singly-exponential in the size of $\Gamma_P$.  Since $\Gamma_P$ is of
linear size in $|\phi|$, this means that $|\autom| \in
2^{\mathcal{O}(|\phi|)}$, \ie the \DFA is of singly exponential size in the
size of the original \DECLARE model. Finally, we solve a reachability
game~\cite{de2007concurrent} on $\autom$ to determine whether Controller
can force the reachability of a final state in $\autom$: if this is the
case, then $\phi$ is realizable, otherwise $\phi$ is unrealizable. Since
reachability games can be solved in polynomial time in the size of
automaton~\cite{de2007concurrent}, this means that the entire algorithm for
\DECLARE realizability works in singly exponential time with respect to the
size of the initial formula.

%% file: sections/5.symbolic.tex

\section{Symbolic Reactive Synthesis for \DECLARE}
\label{sec:symbolic}

In this section, we give a novel algorithm for building an equivalent,
\emph{linear-size}, \emph{symbolic} \DFA starting from (the pastification
of) each \DECLARE model. We then show how to use this automaton to obtain
a symbolic algorithm for \DECLARE realizability. Since symbolic automata
can be exponentially more succinct than the explicit-state ones, this
algorithm is of particular importance for applications.

\subsection{Symbolic \DFAs for pure past \LTLf formulas}
\label{sub:ltlffp:sdfa}

In order to build linear-size, symbolic \DFAs for any \DECLARE model, we
use the following methodology. Given any \DECLARE model $\phi$:
\begin{enumerate*}[label=(\roman*)]
  \item we obtain a linear-size pastification $\pastify(\phi)$ as described
    in \cref{sec:explicit};
  \item we build a linear-size, symbolic \DFA $\autom$ that is equivalent
    to $\pastify(\phi)$.
\end{enumerate*}
In the following, we show a novel translation from general \LTLfFP formulas
(which encompass the pastification of all \DECLARE models) to linear-size
symbolic \DFA.

The technique for the symbolic automata construction that we propose in
this paper is much simpler and efficient then the one
in~\cite{cimatti2021extended,phdgeatti}, since it does not use any counter
bit and uses much less state variables.  For example,
in~\cite{cimatti2021extended} a state variable is used for each subformula
of the initial formula, while here we introduce a state variable only for
the subformulas of the form $\ltl{Y\phi}$ or $\ltl{Z\phi}$.

Let $\phi\in\LTLfFP$ be a formula of pure past \LTLf over the alphabet
$\Sigma$. In the following, we describe how to build a symbolic \DFA
$\autom(\phi) = (X\cup\Sigma,I(X),T(X,\Sigma,\\ X^\prime), F(X,\Sigma))$
such that $\lang(\autom(\phi)) = \lang(\phi)$ and $|\autom(\phi)|$ is linear in
$|\phi|$.

We first define the \emph{closure} of a formula $\phi\in\LTLfFP$.  
\begin{definition}[Closure of \LTLfFP formulas]
  The \emph{closure} of a formula $\phi\in\LTLfFP$ over the alphabet
  $\Sigma$, denoted as $\closure(\phi)$, is the smallest set of formulas
  satisfying the following properties:
  \begin{enumerate}
    \item $\phi \in \closure(\phi)$, and, for each sub-formula $\phi'$ of
      $\phi$, $\phi' \in \closure(\phi)$;
    \item for each $p \in \Sigma$, $p \in \closure(\phi)$ if and only if
      $\neg p \in \closure(\phi)$;
    \item if $\ltl{\phi_1 S \phi_2}$ (resp., $\ltl{\phi_1 T \phi_2}$) is in
      $\closure(\phi)$, then $\ltl{Y(\phi_1 S \phi_2)}$ (resp.,
      $\ltl{wY(\phi_1 T \phi_2)}$) is in $\closure(\phi)$.
  \end{enumerate}
\end{definition}

\mypar{State variables.}
We denote by $\closure_{\ltl{Y}}(\phi)$ (resp.,
$\closure_{\ltl{wY}}(\phi)$) the set of formulas of type $\ltl{Y\phi_1}$
(resp., $\ltl{wY}\phi_1$) in $\closure(\phi)$.
For each formula $\psi$ in
$\closure_{\ltl{Y}}(\phi)\cup\closure_{\ltl{wY}}(\phi)$, we introduce
a state variable $x_{\psi}$ that tracks the truth of
$\psi$, \ie $x_\psi$ holds at instant $i$ of a run $\tau$ iff the automaton
has read a word $\sigma$ such that $\sigma,i \models \psi$.  Formally:
\begin{align}
  X \coloneqq \set{x_\psi \suchthat \psi \in
  \closure_{\ltl{Y}}(\phi)\cup\closure_{\ltl{wY}}(\phi)}
\end{align}

\mypar{Formula for the initial states}
Formula $I(X)$ describes the initial states as the formula
setting all variables in $\closure_{\ltl{Y}}(\phi)$ to false and all
those in $\closure_{\ltl{wY}}(\phi)$ to true:
\begin{align}
  I(X) \coloneqq  \bigwedge_{x \in \closure_{\ltl{Y}}(\phi)} \lnot x \land
                  \bigwedge_{x \in \closure_{\ltl{wY}}(\phi)} x
\end{align}
This forces each formula of type $\ltl{Y\phi_1}$ (resp., of
type $\ltl{wY\phi_1}$) to be false (resp., true) at the first state. Note that there is exactly one satisfying assignment to $I(X)$.

\mypar{Formula for the transition relation}
The formula $T(X,\Sigma,X')$ for the transition relation of $\autom(\phi)$
is defined as the conjunction of the formulas for the transition relation
of each state variable $x \in X$, that we shall define.
We first give the definition of the \emph{stepped normal form} of a given
\LTLfFP formula.

\begin{definition}[Stepped Normal Form]
  Let $\phi$ be a \LTLfFP formula over the alphabet $\Sigma$. Its
  \emph{stepped normal form}, denoted by $\snf(\phi)$, is defined as
  follows:
  \begin{align}
    \snf(\ell) &= \ell 
      \tag*{where $\ell \in \set{p, \neg p}$, for $p \in \Sigma$}                 \\
    \snf(\otimes\,\phi_1) &= \otimes\,\phi_1 
      \tag*{where $\otimes \in \set{\ltl{Y,wY}}$}      \\
    \snf(\phi_1 \otimes \phi_2) &= \snf(\phi_1) \otimes \snf(\phi_2) 
      \tag*{where $\otimes \in \set{\land, \lor}$}       \\
    \snf(\ltl{\phi_1 S \phi_2}) &= \snf(\phi_2) \lor (\snf(\phi_1) \land \ltl{Y(\phi_1 S \phi_2)})  \\
    \snf(\ltl{\phi_1 T \phi_2}) &= \snf(\phi_2) \land (\snf(\phi_1) \lor \ltl{wY(\phi_1 T \phi_2)})
  \end{align}
\end{definition}

We define the $\ground(\cdot)$ function as follows. Given any formula $\phi$
of \LTLfFP, we define $\ground(\phi)$ to be the formula obtained from $\phi$
by replacing each formula of type $\ltl{Y\phi_1}$ (resp., $\ltl{wY\phi_1}$)
with the state variable $x_{\ltl{Y\phi_1}}$ (resp., $x_{\ltl{wY\phi_1}}$).

Let $x_\psi$ be any variable in $X$, where $\psi\coloneqq \otimes(\psi_1)$,
for some $\otimes \in \set{\ltl{Y,wY}}$ and for some
$\psi_1\in\closure(\phi)$. We define the transition relation of
$x_{\otimes(\psi_1)}$ as follows:
\begin{align}
  x^\prime_{\otimes(\psi_1)} \iff \ground(\snf(\psi_1))
\end{align}
Intuitively, the value of $x^\prime_{\otimes(\psi_1)}$ is true at time
point $i+1$ iff $\ground(\snf(\psi_1))$ is true at time point $i$. For every $\psi_1$, the transition relation for $x_{\ltl{Y\psi_1}}$ is
the same as the transition relation for $x_{\ltl{wY\psi_1}}$: the value of
these state variables differs only in the initial state. In
fact, for all the time points different from the first ones, the
\emph{yesterday} and the \emph{weak yesterday} operators have the same
semantics.

\mypar{Formula for final states}
The formula $F(X,\Sigma)$ for the final states captures those states in
which the initial formula $\phi$ holds. We define
$F(X,\Sigma)$ as follows:
\begin{align}
  F(X,\Sigma) \coloneqq \ground(\snf(\phi))
\end{align}
Any trace that reaches, at time point $i$, a state satisfying $F(X,\Sigma)$
is forced to satisfy the original formula $\phi$ at $i$.

Let $\autom(\phi)= (X\cup\Sigma,I(X),T(X,\Sigma,X^\prime), F(X,\Sigma))$.
Crucially, the automaton $\autom(\phi)$ is a symbolic \DFA (recall
\cref{def:sdfa}) because
\begin{enumerate*}[label=(\roman*)]
  \item the formula $I(X)$ has exactly one satisfying assignment;
  \item the formula $T(X,\Sigma,X')$ is of the form $\bigwedge_{x \in
    X}(x^\prime \iff \beta_x(V))$, where $\beta_x(V)$ is a Boolean formula
    over $V (= X \cup \Sigma)$, for each $x \in X$.
\end{enumerate*}
The following theorem establishes the correctness of the procedure for
building linear-size, symbolic \DFAs starting from any \DECLARE model.

\begin{restatable}{theorem}{thsdfacorrect}
\label{th:sdfa:correct}
  For any \DECLARE model $\phi$, the automaton $\autom(\phi)$ is such
  that:
  \begin{enumerate*}[label=(\roman*)]
    \item $\autom(\phi)$ is symbolic \DFA;
    \item $\lang(\autom(\phi)) = \lang(\phi)$;
    \item $|\autom(\phi)| \in \mathcal{O}(|\phi|)$.
  \end{enumerate*}
\end{restatable}
\begin{proof}
  We begin proving the first and the third point.

  Firstly, the formula $I(X)$ for the initial states has exactly one
  satisfying assignment, \ie the one that sets to true (resp., to false)
  all the state variables corresponding to a formula in the closure of type
  $\ltl{Z\phi_1}$ (resp., of type $\ltl{Y\phi_1}$).
  Moreover, the formula $T(X,\Sigma,X')$ for the transition function is of
  the form $\bigwedge_{x\in X} x' \iff \beta(X,\Sigma)$, where
  $\beta(X,\Sigma)$ is a Boolean formula over the variables in $X$ and
  $\Sigma$.
  This implies that $\autom(\phi)$ is a symbolic \DFA. 

  It is easy to see that the number of state variables is \emph{linear} in
  the size of $\phi$, \ie $|X|\in\mathcal{O}(|\phi|)$. Moreover, the
  formulas $I(X)$, $T(X,\Sigma,X')$ and $F(X)$ are of size linear in $|X|$
  and thus they are linear in $|\phi|$ as well.
  This means that $|\autom(\phi)| \in \mathcal{O}(|\phi|)$.

  We now show that $\lang(\autom(\phi)) = \lang(\phi)$.
  We first prove that $\lang(\phi) \subseteq \lang(\autom(\phi))$. Let
  $\sigma=\seq{\sigma_0,\sigma_1,\dots,\sigma_n}$ be a word in
  $\lang(\phi)$. We show that $\sigma\in\lang(\autom(\phi))$.
  Let $\tau=\seq{X_0,X_1,\dots,X_n}$ be the sequence in $(2^X)^*$ defined
  as follows:
  \begin{itemize}
    \item $X_0 = \set{x_{\ltl{Z\phi_1}} \suchthat \ltl{Z\phi_1} \in
      \closure(\phi)}$;
    \item $X_{i+1} = \set{ x_{\ltl{Z\phi_1}}, x_{\ltl{Y\phi_1}} \suchthat
      (X_i,\sigma_i) \models \ground(\snf(\phi_1)), 
      \ltl{Z\phi_1} \in \closure(\phi),
      \ltl{Y\phi_1} \in \closure(\phi)
      }$.
  \end{itemize}
  We define $\pi$ as the sequence
  $\seq{(X_0,\sigma_0),(X_1,\sigma_1),\dots,(X_n,\sigma_n)}$.
  By definition of $\autom(\phi)$, it holds that $\pi$ is a run of
  $\autom(\phi)$ induced by $\sigma$. We now show that $\pi$ is
  \emph{accepting} as well.

  By induction on the structure of $\phi$, it is easy to prove that, for
  all $\psi\in\closure(\phi)$ and for all $0\le i\le n$, it holds that:
  \begin{align}
    \seq{\sigma_0,\dots,\sigma_i} \models \psi
    \Iff
    (X_i,\sigma_i) \models \ground(\snf(\psi))
  \end{align}
  Now, since by hypothesis $\seq{\sigma_0,\dots,\sigma_n} \models \phi$, we
  have that $(X_n,\sigma_n) \models \ground(\snf(\phi))$, that is,
  $(X_n,\sigma_n) \models F(X,\Sigma)$.
  This means that $\pi$ is an accepting run of $\autom(\phi)$ and thus
  $\sigma\in\lang(\autom(\phi))$.

  We now show that $\lang(\autom(\phi)) \subseteq \lang(\phi)$. Let
  $\sigma=\seq{\sigma_0,\dots,\sigma_n}$ be a word in
  $\lang(\autom(\phi))$. We show that $\sigma\in\lang(\phi)$.
  Since $\sigma$ is accepted by $\autom(\phi)$, there exists an
  \emph{accepting} run $\pi=\seq{(X_0,\sigma_0),\dots,(X_n,\sigma_n)}$.
  Like above, for all $\psi\in\closure(\phi)$ and for all $0\le i\le n$, it
  holds that:
  \begin{align}
    \seq{\sigma_0,\dots,\sigma_i} \models \psi
    \Iff
    (X_i,\sigma_i) \models \ground(\snf(\psi))
  \end{align}
  Since by hypothesis, $(X_n,\sigma_n) \models \ground(\snf(\phi))$, it
  holds that $\seq{\sigma_0,\dots,\sigma_n} \models \phi$, \ie
  $\sigma\in\lang(\phi)$.
  \qed
\end{proof}

\subsection{A symbolic algorithm for \DECLARE realizability}
\label{sub:symbolic:alg}

The construction of linear-size, symbolic \DFAs starting from any \LTLfFP
formula, together with the pastification of \DECLARE shown in
\cref{sec:explicit}, make it possible to devise a symbolic algorithm for
solving \DECLARE realizability. 

The algorithm works as follows. Let $\phi_E$ and $\phi_C$ be two \DECLARE
models for the assumptions on the environment and the guarantees for the
controller.
Let $\Gamma$ be the formula $\mathtt{simple}_{\mathtt{Con}}(\Cset) \land
((\mathtt{simple}_{\mathtt{Env}}(\Uset) \land \phi_E) \to \phi_C)$. Recall
that, by \cref{lemma:simple:general:real}, $\Gamma$ is an \LTLf formula and
is realizable (over general traces) iff $(\phi_E,\phi_C)$ is realizable.
As the first step, starting from the four formulas inside $\Gamma$ (see
\cref{cor:past:correctness}), the algorithm produces a linear-size
pastification $\Gamma_P$ for $\Gamma$.
Then, it builds a linear-size, symbolic \DFA $\autom$ such that
$\lang(\autom)=\lang(\Gamma_P)$, as shown in this section. As the last
step, the algorithm solves a \emph{symbolic reachability game} over
$\autom$: the aim of Controller is to choose a sequence of values for the
controllable variables so that, no matter what values for the
uncontrollable variables are chosen by Environment, the trace induced by
the play in $\autom$ is \emph{accepting}, \ie, it eventually visits
a final state of $\autom$.  Since by construction $\autom$ recognizes
exactly the language of $\phi$, the play satisfies $\phi$, and thus
Controller has a winning strategy for $\phi$.
Reachability games can be solved efficiently in a fully symbolic fashion
over a symbolic \DFA by computing the \emph{strong preimage} of the
transition function of the automaton, until a fixpoint is reached. We refer
to~\cite{de2007concurrent,jacobs2017first} for more details.  Moreover,
since the organization of the SYNTCOMP~\cite{jacobs2017first}, many
optimized tools have been proposed to solve reachability
games over symbolic arenas. Hence, \DECLARE realizability is
likely to be solved very efficiently in practice.

%% file: sections/6.conclusions.tex

\section{Conclusions}
\label{sec:conclusions}

We have introduced the realizability and reactive synthesis
problem for temporal specifications written in \DECLARE (a well-established pattern-based language for specifying declarative, constraint-based business processes), whose semantics is defined through \LTL interpreted over simple finite traces.
We have shown how the problem can be recasted as \LTLf realizability.
However, the doubly-exponential time complexity of realizability
 for \LTLf suggests that this approach may be not optimal for
\DECLARE, strict subsumed by \LTLf.  In fact, we give
a singly exponential time algorithm for \DECLARE realizability, based on
the technique of pastification, allowing to leverage the singly
exponential \DFA construction starting from pure past  \LTLf formulas,
proving that \DECLARE is a fragment of \LTLf with an efficient synthesis
problem.
We also give a symbolic version of the previous algorithm, which
 has the potential of being applied in practice.

The study of the optimality of our algorithm in terms of
worst-case complexity surely deserves attention.
An interesting future development is the implementation of the symbolic
algorithm and its experimental evaluation, which in turn requires to
collect related benchmarks.
There are a lot of other tasks in formal verification, other than
synthesis, in which deterministic (and symbolic) automata play an important role. 
For example, we plan to apply the proposed symbolic automata
construction for \DECLARE for monitoring problems, in which the construction of a deterministic, symbolic monitor
starting from a \DECLARE specification can be done more efficiently. 
More to that, the obtained results pave the way towards computational improvements in other automata-based tasks from the repertoire  of techniques available for \DECLARE~\cite{CiccioM22,DDMM22}.